\documentclass[10pt]{iopart}

\expandafter\let\csname equation*\endcsname\relax
\expandafter\let\csname endequation*\endcsname\relax
\usepackage{amsmath,amssymb}
\usepackage[bookmarks=true,hyperfigures=true,colorlinks=true,linkcolor=black,
citecolor=black,urlcolor=black,bookmarksnumbered,hidelinks]{hyperref}
\usepackage{hyperref}
\usepackage{cleveref}
\usepackage{cite}
\usepackage{xcolor}

\newcommand{\la}{\lambda}
\newcommand{\op}{\operatorname}
\newcommand{\wt}{\widetilde}
\newcommand{\bd}{\boldsymbol}
\newcommand\beq{\begin{equation}}
\newcommand\eeq{\end{equation}}
\newcommand\bea{\begin{align}}
\newcommand\eea{\end{align}}
\newcommand\bed{\begin{aligned}}
\newcommand\eed{\end{aligned}}
\newcommand\bear{\begin{array}}
\newcommand\eear{\end{array}}
\newcommand\ds{\displaystyle}
\numberwithin{equation}{section}

\begin{document}
\title[Triangular solutions to the reflection equation for $U_q(\widehat{sl_n})$]
{Triangular solutions to the reflection equation for $U_q(\widehat{sl_n})$}
\author{Dmitry Kolyaskin and Vladimir V Mangazeev}
\address{Department of Theoretical \& Fundamental Physics, Research School of Physics,
Australian National University, Canberra, ACT 2601, Australia}
\ead{dmitrii.koliaskin@anu.edu.au}
\ead{vladimir.mangazeev@anu.edu.au}
\vspace{10pt}

\begin{abstract}
We study solutions of the reflection equation related to
the quantum affine algebra $U_q(\widehat{sl_n})$.
First, we explain how to construct a family of stochastic
integrable vertex models with fixed boundary conditions.
Then, we construct upper- and lower-triangular solutions of the reflection equation
related to
symmetric tensor representations of $U_q(\widehat{sl_n})$ with arbitrary spin.
We also prove the star-star relation
for the Boltzmann weights of the Ising-type model, conjectured by Bazhanov and Sergeev,
and use it to verify certain properties of the solutions obtained.
\end{abstract}
\vspace{1pc}

\maketitle

\section{Introduction}

Stochastic integrable vertex models play an important role in the analysis of $1+1$-dimensional systems out-of-equilibrium
\cite{GS92,BorCG2016,BP16},
interacting particle systems with exclusion \cite{Sp70,DEHP93,Der98,Sas99,DGE05}, and
KPZ universality \cite{KPZ86,Cor12}. From one side, they provide interesting examples of interacting
Markovian systems; from the other side, one can analyse their critical behavior and explicitly calculate  some
physical quantities like correlation functions.

Corwin and Petrov in \cite{Cor2015}  (see also \cite{BP2018} for a review)  showed that many currently known
integrable $1+1$-dim KPZ models
come from suitable limits of stochastic higher-spin vertex models.
Originally, they studied these higher-spin models  on the line (see also \cite{Povol13,KMMO16}).
However, later, this was generalized to half-line, half-quadrant processes with open
and special fixed boundary conditions (see, for example, \cite{BBCW2018,BP2018,gmw2023,gg23}).

Sklyanin \cite{Sklyanin:1988yz} developed a general theory of quantum integrable systems with boundaries
based on the so-called reflection equation. The reflection equation
first appeared in a 2-dim factorised scattering theory  on a half-line \cite{Cherednik}, with
the scattering matrix satisfying the Yang-Baxter equation \cite{Yang:1968,Zam79,Bax72,Bax82}.

In this paper, we construct special triangular solutions of the reflection equation for the higher-spin stochastic
vertex models related to symmetric tensor representations of $U_q(\widehat{sl_n})$.
These solutions generalise some of our results from \cite{Man2019}. Let us also mention a recent paper 
\cite{Frassek22}, where the author constructed a special one-parametric family of integrable open boundaries for 
the so-called $q$-Hahn processes,
related to $U_q(\widehat{sl_2})$.

It is important to note that
we work with a {\it stochastic} $R$-matrix, which is a twisted version \cite{Perk:1981nb} of
the standard $U_q(\widehat{sl_n})$ $R$-matrix.
As was noticed in \cite{BM2016}, this leads to a factorised $L$-operator, which may be
why all our solutions possess factorised matrix elements.

The paper is organised as follows. In Section \ref{Section 2} we review the theory of boundary integrable systems
and show how to construct stochastic versions of the transfer-matrix and hamiltonian.
In Section \ref{Section 3} we introduce required notations, define the stochastic $U_q(\widehat{sl_n})$ $R$-matrix
and discuss its properties. We also introduce two higher-spin $L$-operators required later in the text.
In Section \ref{Section 4} we prove the star-star relation from \cite{BS2023}, which we use later.
Sections \ref{Section 5} and \ref{Section 6} contain the main results of the paper. Namely, we construct
left and right boundary solutions for the reflection equation in (\ref{Upper-triangular solution}),
  (\ref{K-matrix, lower-triangular solution}), (\ref{Left boundary}) and (\ref{Left boundary, lower-triangular solution})
  and prove their properties.
In Section \ref{Section 7} we give the proof of the reflection equation for a generalized model presented in \cite{Man2019}.
In Conclusion, we discuss the  results obtained and unsolved problems.
Finally, in Appendices, we prove some technical results used in the text.

\section{Reflection equation and commuting transfer-matrices}\label{Section 2}
We start with the Yang-Baxter equation
\beq\label{YBE1}
    R_{12}\left(\frac{x}{y}\right) R_{13}\left(\frac{x}{z}\right) R_{23}\left(\frac{y}{z}\right)=
    R_{23}\left(\frac{y}{z}\right) R_{13}\left(\frac{x}{z}\right) R_{12}\left(\frac{x}{y}\right),
\eeq
whose solutions $ R_{12}\left(x\right)\in\text{End}\left(V_1\otimes V_2\right)$
act non-trivially in the tensor product of two
vector spaces and describe the behaviour of the system in the bulk.
We assume that the $R$-matrix satisfies the following properties:
\begin{itemize}
    \item $ R^{t_2}_{12}\left(x\right)$ is non-degenerate, i.e. one can correctly define the matrix
    \beq\label{R-tilde}
    \wt R_{12}\left(x\right)=\left(\left(R_{12}^{t_2}\left(x\right)\right)^{-1}\right)^{t_2},
    \eeq
\end{itemize}
{where $t_i$ denotes transposition with respect to $V_i$, i.e. in the matrix components 
it has the form $\left[R_{12}^{t_2}\left(x\right)\right]_{i, j}^{i^{\prime}, j^{\prime}}=
\left[R_{12}\left(x\right)\right]_{i, j^{\prime}}^{i^{\prime}, j}.$}
\begin{itemize}
    \item  Regularity condition
        \beq\label{YBE2}
        R_{1 2}\left(1\right) = P_{12},
    \eeq
\end{itemize}
    where we assume that $V_1\cong V_2\equiv V$. Here $P$ is a permutation operator,
    acting in a usual way $P(v \otimes w)=w \otimes v$, where $v,w\in V$.
\begin{itemize}
    \item Unitarity condition
    \beq\label{YBE3}
        R_{12}(x) R_{21}\left(x^{-1}\right)=f(x) I \otimes I,
    \eeq
\end{itemize}
     where $f(x)$ is a scalar function satisfying $f(x)=f(x^{-1})$. In fact, one can check that if $V_1\cong V_2$,
     then the last property directly follows from the regularity condition (\ref{YBE2}) and the Yang-Baxter equation (\ref{YBE1})
     (see, for example, \cite{Bazhanov87}).
\begin{itemize}
    \item Crossing unitarity condition
        \beq\label{YBE4}
        M_1 R_{12}^{t_1}(x) M_1^{-1} R_{21}^{t_1}\left(\left(r x\right)^{-1}\right)=g(x) I \otimes I,
    \eeq
\end{itemize}
    where $r\in \mathbb{C}\setminus\{0\}$, and $M\in \text{End}(V)$ is a constant matrix. Note that the crossing unitarity
    is a stronger condition than (\ref{R-tilde}) \cite{DANCER2009456, Man2019}, since the expression
    \beq\label{YBE5}
         \wt R_{21}\left(x\right)=\dfrac{1}{g(1/(r x))}\left(M^{t_1}_1\right)^{-1}R_{12}\left(1/(r x)\right)M^{t_1}_1
    \eeq
    directly follows from (\ref{YBE4}). In particular, if we consider the $R$-matrix as an intertwiner of
    finite-dimensional $U_q(\widehat{\mathfrak{g}})$ modules, then it follows from the quantum group arguments
    that conditions of unitarity \eqref{YBE3} and crossing unitarity (\ref{YBE4}) are satisfied \cite{FR1992,Vlaar2015}.
    In this instance, $M=q^{-2\rho}$, $r=q^{h^{\vee}},$ where $\rho$ is a half-sum of positive roots, and $h^{\vee}$
    is a dual Coxeter number of the Lie algebra $\mathfrak{g}.$ In what follows, we will focus on
    $\mathfrak{g}=sl_n(\mathbb{C})$ case, and as we will see in the next chapter, the matrix $M$ is diagonal and
    satisfies an additional property
\beq\label{YBE6}
    \left[M \otimes M, R_{12}(x)\right]=0.
\eeq

Now we come to describing the boundaries: for a given solution to the Yang-Baxter equation (\ref{YBE1}),
we define the so-called reflection equation
\beq\label{YBE7}
    R_{12}\left(\frac{x}{y}\right) K_1\left(x\right) R_{21}\left(x y\right) K_2\left(y\right)=K_2\left(y\right) R_{12}
    \left(x y\right) K_1\left(x\right) R_{21}\left(\frac{x}{y}\right)
\eeq
and the dual reflection equation
\beq\label{YBE8}
   \wt{K}_2\left(y\right)\wt R_{21}(x y) \wt{K}_1\left(x\right) R_{21}\left(\frac{y}{x}\right)=R_{12}\left(\frac{y}{x}\right)
   \wt{K}_1\left(x\right)\wt R_{12}(x y) \wt{K}_2\left(y\right),
\eeq
where the matrix $\wt R_{12}(x)$ is defined in (\ref{R-tilde}). The solutions $K_i\left(x\right), \wt{K}_i\left(x\right)$
act non-trivially on $V_i$ $(i=1,2)$ and describe the right and left boundaries, respectively. Knowing the solution
to the reflection equation (\ref{YBE7}) and using the commutativity property (\ref{YBE6}), the solution to
the dual reflection equation (\ref{YBE8}) can be expressed in the form
\beq\label{YBE9}
     \wt{K}_1(x)=M^{-1}K\left(\dfrac{1}{r^{1/2}x}\right).
\eeq
Following \cite{Sklyanin:1988yz,Crampe_2014}, we can also consider another reflection equation related to
the inverse $R$-matrix $R_{21}\left(x^{-1}\right)\cong R_{12}^{-1}\left(x\right)$:
\beq \label{YBE10}
R_{21}\left(\frac{y}{x}\right) \bar{K}_1\left(x\right) R_{12}\left(\frac{1}{x y}\right) \bar{K}_2\left(y\right)=
\bar{K}_2\left(y\right) R_{21}\left(\frac{1}{x y}\right) \bar{K}_1\left(x\right) R_{12}\left(\frac{y}{x}\right).
\eeq
It is stated that if $\bar{K}(x)$ is a solution to equation ($\ref{YBE10}$), then the expression
\beq\label{YBE11}
    \wt{K}_1(x)=\op{tr}_0\left(\bar{K}_0\left(\dfrac{1}{x}\right)\wt R_{01}\left(x^2\right) P_{01}\right)
\eeq
solves the dual reflection equation (\ref{YBE8}). The proof of this fact is quite technical, and
we provide it in \ref{AppA} for the reader's convenience.
The mapping (\ref{YBE11}) defines an isomorphism between the solutions to
the dual reflection equation (\ref{YBE8}) and
the equation (\ref{YBE10}) (see \cite{Sklyanin:1988yz} for this and other isomorphisms):
the inverse mapping has the form
\beq\label{YBE12}
    \bar{K}_1(x)=\op{tr}_0\left(\wt{K}_0\left(\dfrac{1}{x}\right)R_{01}\left(\dfrac{1}{x^2}\right) P_{01}\right),
\eeq
which we also prove in \ref{AppA}.

Substituting $y=1$ into the reflection equation \eqref{YBE7}, we get
\beq
[R_{12}(x)K_1(x)R_{21}(x), K_2(1)]=0 \label{YBE12a}
\eeq
which should be true for any $x$. Apparently, the only reasonable solution to \eqref{YBE12a} is $K(1)\sim I$.
In this paper we shall always assume the normalization
\beq
K(1)=\it{I}.\label{YBE12b}
\eeq

Solutions to the reflection equation (\ref{YBE7}) and the dual reflection equation (\ref{YBE8})
allow us to construct a double-row transfer matrix
\beq \label{YBE13}
T(x)=
\op{tr}_0\left[\wt{K}_0\left(x\right)R_{01}\left(x\right)\dots R_{0N}\left(x\right)K_0
\left(x\right)R_{N0}\left(x\right)\dots R_{10}\left(x\right) \right].
\eeq

Transfer-matrices \eqref{YBE13} form a one-parameter family of mutually commuting quantities:
\beq\label{YBE14}
     \left[T(x),T(y)\right]=0,
\eeq
and this property implies the integrability of the model. The proof of \eqref{YBE14} can be found in \cite{Sklyanin:1988yz}.

Taking into account \eqref{YBE2} and \eqref{YBE12b}, we obtain
\beq\label{YBE13a}
T(1)=\op{tr}{\bigl[}\wt{K}\left(1\right){\bigr]}\,\bd{\it{I}}.
\eeq

Finally, we can calculate the Hamiltonian of the integrable system associated with the transfer-matrix \eqref{YBE13}
\begin{equation} \label{YBE15}
H =-\dfrac{1}{4}\dfrac{\partial}{\partial x}\ln T(x)\bigg|_{x=1}=B_{L}+\sum\limits_{k=1}^{N-1}H_{k,k+1}+B_{R},
\end{equation}
where
\beq\label{YBE16}
B_L=\dfrac{1}{4}\dfrac{\bar{K}^{\prime}_1\left(1\right)}{\operatorname{tr}\widetilde{K}(1)},
\quad
H_{k,k+1}=-\dfrac{1}{2}R^{\prime}_{k,k+1}(1)P_{k,k+1},\quad
B_R=-\dfrac{1}{4}{K}^{\prime}_N\left(1\right).
\eeq
The operators $B_L$ and $B_R$ act only in the first and the $N$-th component of the Hilbert space. Let us notice that
the operator $B_L$ at the left boundary is proportional to the derivative of the matrix ${\bar{K}}(x)$
 as can be easily seen from differentiating \eqref{YBE12}.

Now let us discuss under what conditions the Hamiltonian and the transfer-matrix define stochastic processes.

We say that an $M\times M$ matrix $A_{ij}$ is (left) stochastic if it satisfies the condition
\beq\label{YBE17}
\sum_{i=1}^M A_{ij}=1  \mbox{ for any }j \>\mbox{ or }\> \langle\bd{1}|A=\langle\bd{1}|, \quad \langle\bd{1}|=\langle1\ldots1|,
\eeq
that is, $\langle\bd{1}|$ is a row vector whose components are all equal to 1.

If the transfer-matrix $T(x)$ \eqref{YBE13} is left stochastic,
\beq\label{YBE18}
\langle\bd{1}|\otimes\ldots\otimes\langle\bd{1}|T(x)=\langle\bd{1}|\otimes\ldots\otimes\langle\bd{1}|,
\eeq
and matrix elements of $T(x)$ are positive
then $T^t(x)$ defines a discrete-time Markov process in a quantum space $V\otimes\ldots\otimes V$. The corresponding Hamiltonian
\eqref{YBE15} satisfies the property
\beq\label{YBE19}
\langle\bd{1}|\otimes\ldots\otimes\langle\bd{1}|H=0
\eeq
and the continuous-time Markov generator is defined by $M=-H^t$. The inverse statement is harder to establish and it is not always true.

First, we call matrices $R_{12}(x)$, $K(x)$ and $\bar{K}(x)$ stochastic if they are left stochastic
\beq\label{YBE20}
\langle\bd{1}|\otimes\langle\bd{1}|R_{12}(x)=\langle\bd{1}|\otimes\langle\bd{1}|,\quad
\langle\bd{1}|K(x)=\langle\bd{1}|, \quad \langle\bd{1}|\bar{K}(x)=\langle\bd{1}|.
\eeq
Differentiating \eqref{YBE20} with respect to $x$ and using (\ref{YBE15}-\ref{YBE16}), we conclude that
\beq\label{YBE21}
\langle\bd{1}|\otimes\ldots\otimes\langle\bd{1}|H=0.
\eeq
Now let us remember that the Hamiltonian \eqref{YBE15} commutes with the transfer-matrix
\beq
\label{YBE21a}
[T(x),H]=T(x)\,H -H\, T(x)=0.
\eeq
Applying $\langle\bd{1}|\otimes\ldots\otimes\langle\bd{1}|$ from the left, we obtain
\beq\label{YBE22}
\langle\bd{1}|\otimes\ldots\otimes\langle\bd{1}|T(x)H=\langle\bd{1}|\otimes\ldots\otimes\langle\bd{1}|H\,T(x)=0.
\eeq
If the rank of the matrix $H$ is $\mbox{dim}(V)^N-1$, then $\langle\bd{1}|\otimes\ldots\otimes\langle\bd{1}|$
is the only left zero vector of $H$ and we conclude that
\beq\label{YBE23}
\langle\bd{1}|\otimes\ldots\otimes\langle\bd{1}|T(x)=c(x)\langle\bd{1}|\otimes\ldots\otimes\langle\bd{1}|.
\eeq
Dividing $T(x)$ by a scalar factor $c(x)$, we obtain the condition \eqref{YBE18}. If the rank of $H$ is smaller than
$\mbox{dim}(V)^N-1$, we cannot say anything about the stochasticity of $T(x)$. Indeed, in the case of periodic boundary
conditions the hamiltonian may have additional zero modes and the transfer matrix is not stochastic in general.
However, in all cases with fixed boundary conditions, which we considered, the rank of the Hamiltonian $H$ was $\mbox{dim}(V)^N-1$
 and this implies \eqref{YBE23}.

Let us also notice that we don't know how to show the stochasticity of the transfer-matrix \eqref{YBE13} directly from
\eqref{YBE20}. The main difficulty is that $T(x)$ is defined in terms of $\wt{K}(x)$ related to $\bar{K}(x)$ via \eqref{YBE11}.

\section{Stochastic $R$-matrix for  \texorpdfstring{$U_q(\widehat{sl_n})$}{Uq(sln)}}\label{Section 3}
In this section we introduce the stochastic $R$-matrix
related to the symmetric tensor representations of the quantum affine algebra $U_q(\widehat{sl_n})$ and consider its basic properties.

First, we start by introducing index notations. We denote by $\bd{i}=\{i_1, \dots, i_m\}$, the $m$-component set of non-negative integers
$i_k\in\mathbb{Z}_{+},$ $k=1,\dots m$. We also define some operations on these sets:
\beq\label{operations on vectors}
    |\bd{a}|\stackrel{\text{def}}{=}\sum_{k=1}^{m}a_k, \qquad (\bd{a}, \bd{b})\stackrel{\text{def}}{=}\sum\limits_{k=1}^{m} a_k b_k,
    \qquad  Q(\bd{a},\bd{b})\stackrel{\text{def}}{=}\sum_{1 \leq l<k\leq m}a_l b_k,
\eeq
where $\bd{a}$, $\bd{b}\in\mathbb{Z}_{+}^{m}$. The permutation $\tau$ is defined as
    \beq\label{tau}
        \tau\left\{i_1, \ldots, i_m\right\}=\left\{i_m, i_{m-1}, \ldots, i_1\right\}.
    \eeq
We will also need one more operation on indices: for a given $J\in \mathbb{N}$ and  $\bd{i}\in\mathbb{Z}_{+}^{m}$, $|\bd{i}|\le J$,
  we introduce the mapping
 \beq\label{sigma}
        \sigma\left\{i_1, \ldots, i_{m}\right\}=\left\{i_2, i_{3}, \ldots, i_{m}, J-|\bd{i}|\right\}.
\eeq

For a positive integer $J\in\mathbb{N}$ and $n\ge 2$, we define a finite-dimensional vector space $V_J^{(n)}$
with basis vectors having the form
\beq\label{Basis vectors}
    |\bd{i}\rangle=|i_1,i_2,\dots,i_{n-1}\rangle, \quad |\bd{i}|\le J,
\eeq
 (do not confuse this notation with the row vectors from the previous section). The dimension of the  space $V_J^{(n)}$ is $\binom{J+n-1}{n-1}$.
We will also use an equivalent notation for the basis vectors of $V_J^{(n)}$:
\beq\label{Basis vectors, equivalent notations}
    |\bar{\bd{i}}\rangle=|i_1,i_2,\dots,i_{n-1},i_n\rangle, \quad |\bar{\bd{i}}|=J,
\eeq
where we extended the vector $|\bd{i}\rangle$ by adding an additional component $i_n=J-|\bd{i}|$.
The introduced vector space naturally inherits the structure of finite-dimensional symmetric tensor representations
of the quantum affine algebra $U_q(\widehat{sl_n})$ \cite{KMMO16}. For further purposes, it will be essential for us
to consider a realization of the symmetric tensor representation of $sl_n(\mathbb{C})$ Lie algebra on $V_J^{(n)}$:
    \beq\label{sln representation}
    e_i|\bd{\bar{\alpha}}\rangle=\alpha_{i+1}|\bd{\bar{\alpha}}+\bd{e}_{i}-\bd{e}_{i+1}\rangle,
    \quad f_i|\bd{\bar{\alpha}}\rangle=\alpha_{i}|\bd{\bar{\alpha}}-\bd{e}_{i}+\bd{e}_{i+1}\rangle,
    \quad h_i|\bd{\bar{\alpha}}\rangle=(\alpha_{i}-\alpha_{i+1})|\bd{\bar{\alpha}}\rangle,
    \eeq
where $e_i$, $f_i$ and $h_i$ ($i=1,\dots n-1$) are the standard $sl_n(\mathbb{C})$ generators,
$|\bd{\bar{\alpha}}\rangle=|\alpha_1, \alpha_2, \dots, \alpha_{n}\rangle$
as defined in (\ref{Basis vectors, equivalent notations}) and $\bd{e}_j$ represents the vector with $1$
at the $j$-th place and 0 otherwise. For what follows, we will need to find an expression for
the half-sum of positive roots $\rho$ of $sl_n(\mathbb{C})$, and a simple analysis shows that it has the form
\beq\label{Sum of positive roots}
    2\rho|\bd{\bar{\alpha}}\rangle=\left(-2\sum\limits_{k=1}^{n-1}(n-k)\alpha_k+(n-1)J\right)|\bd{\bar{\alpha}}\rangle.
\eeq

Now we discuss the stochastic $R$-matrix related to symmetric tensor representations,
see \cite{KMMO16, BM2016} for more details. For arbitrary positive integers $I, J\in\mathbb{N}$ we introduce a linear operator
$S_{I, J}(x)\in\op{End}(V^{(n)}_I\otimes V^{(n)}_J)$, whose action on the basis
$\left|\bd{i}^{\prime}\right\rangle \otimes\left|\bd{j}^{\prime}\right\rangle$ (in terms of (\ref{Basis vectors})) has the form
\beq
    S_{I, J}(x)\left|\bd{i}^{\prime}\right\rangle \otimes\left|\bd{j}^{\prime}\right\rangle=
    \sum_{\bd{i}, \bd{j}} \left[S_{I, J}(x)\right]_{\bd{i}, \bd{j}}^{\bd{i}^{\prime}, \bd{j}^{\prime}}|\bd{i}\rangle \otimes|\bd{j}\rangle.
\eeq
The matrix elements can be written in a compact way as
\begin{align}\label{R-matrix}
    &\left[S_{I, J}(x)\right]_{\boldsymbol{i}, \boldsymbol{j}}^{\boldsymbol{i}^{\prime},
    \boldsymbol{j}^{\prime}}=\nonumber\\
    &=\delta_{\boldsymbol{i}+\boldsymbol{j},
    \boldsymbol{i}^{\prime}+\boldsymbol{j}^{\prime}} \sum_{\boldsymbol{m}+\boldsymbol{n}=
    \boldsymbol{i}+\boldsymbol{j}} \Phi_{q^2}\left(\boldsymbol{m}-\boldsymbol{j} \mid \boldsymbol{m} ;
    \frac{q^{J-I}}{x^2}, \frac{q^{-I-J}}{x^2}\right) \Phi_{q^2}\left(\boldsymbol{n}
    \mid \boldsymbol{j}^{\prime} ; \frac{x^2}{q^{I+J}}, q^{-2 J}\right),
\end{align}
where we defined the function
\beq\label{phi-function} 
\bed
\Phi_q(\bd{\gamma} \mid \bd{\beta} ; \la, \mu) & =q^{Q(\bd{\beta}-\bd{\gamma},\bd{\gamma})}
\left(\frac{\mu}{\la}\right)^{|\bd{\gamma}|} \frac{(\la ; q)_{|\bd{\gamma}|}
\left(\frac{\mu}{\la} ; q\right)_{|\bd{\beta}|-|\bd{\gamma}|}}{(\mu ; q)_{|\bd{\beta}|}} \prod_{s=1}^{n-1}\left[\bear{l}
\beta_s \\
\gamma_s
\eear\right]_q
\eed
\eeq
for $\bd{\beta}$, $\bd{\gamma}\in\mathbb{Z}_{+}^{n-1}$ and $\la$, $\mu\in\mathbb{C}$, first introduced in \cite{KMMO16}.
Here we use the standard notation for $q$-Pochhammer symbols
\beq
    (x ; q)_n\stackrel{\text{def}}{=}\begin{cases}\prod\limits_{k=0}^{n-1}\left(1-x q^k\right), & n>0, \\
    1, & n=0, \\ \prod\limits_{k=0}^{-n-1}\left(1-x q^{n+k}\right)^{-1}, & n<0,\end{cases}
\eeq
and $q$-binomial coefficients
\beq
    \left[\bear{l}
m \\
k
\eear\right]_q\stackrel{\text{def}}{=}\frac{(q ; q)_m}{(q ; q)_k(q ; q)_{m-k}}.
\eeq

Let us notice that the original parameter $q$ of the quantum group $U_q(\widehat{sl_n})$ enters the expression for
the stochastic $R$-matrix only via $q^2$ as indicated by the subscript of the function $\Phi_{q^2}$ in \eqref{R-matrix}. 
Such notations were introduced in \cite{KMMO16} and we prefer to follow them.

The introduced $R$-matrix was constructed in \cite{BM2016} from the corresponding 3D $R$-operators, where it was shown that
it satisfies the Yang-Baxter equation defined on $V^{(n)}_I\otimes V^{(n)}_J \otimes V^{(n)}_K$
\beq\label{Yang-Baxter for stochastic R-matrix}
    S_{I,J}\left(\frac{x}{y}\right) S_{I,K}\left(\frac{x}{z}\right) S_{J,K}\left(\frac{y}{z}\right)=
    S_{J,K}\left(\frac{y}{z}\right) S_{I,K}\left(\frac{x}{z}\right) S_{I,J}\left(\frac{x}{y}\right).
\eeq
We can also obtain different symmetries of the $R$-matrix (\ref{R-matrix}) from the symmetries of corresponding 3D $R$-matrix.
Before we describe them, we define a symmetric version of the $R$-matrix (\ref{R-matrix})
\beq\label{Symmetric R-matrix}
\left[\bar{R}_{I, J}(x)\right]_{\bd{i}, \bd{j}}^{\bd{i}^{\prime}, \bd{j}^{\prime}}=
x^{|\bd{i}|-|\bd{i}^{\prime}|}q^{\left(\bd{i}^{\prime},\bd{j}^{\prime} \right)-\left(\bd{i},\bd{j}
\right)+J|\bd{i}|-I|\bd{j}^{\prime}|-Q(\bd{j}, \bd{i})+Q(\bd{i}^{\prime}, \bd{j}^{\prime})}
\left[S_{I, J}(x)\right]_{\bd{i}, \bd{j}}^{\bd{i}^{\prime}, \bd{j}^{\prime}}.
\eeq
One can check that in the case  $n=2$ and $I=J=1$, the  $R$-matrix \eqref{R-matrix} corresponds
to the stochastic symmetric 6-vertex model
\beq\label{6-vertex}
S_{12}(x)=\left(
\begin{array}{cccc} 1 & 0 & 0 & 0\\
0 & \frac{q^2(x^2-1)}{q^2x^2-1} & \frac{(q^2-1)x^2}{q^2x^2-1} & 0\\
0 & \frac{(q^2-1)}{q^2x^2-1} & \frac{(x^2-1)}{q^2x^2-1} & 0\\
0 & 0 & 0 & 1\\
\end{array}\right).
\eeq

It will  be also convenient to represent the function $\Phi$ from
(\ref{phi-function}) in another form
\beq\label{phi-function alternative}
    \Phi_{q^2}(\bd{\gamma} \mid \bd{\beta} ; \la^2, \mu^2)=\dfrac{\mu^{|\bd{\gamma}|}}
    {\la^{|\bd{\beta}|}}\dfrac{V_{\la}(\bd{\gamma},0)V_{\mu / \la}(\bd{\beta},\bd{\gamma})}{V_{\mu}(\bd{\beta},0)},
\eeq
where the function $V_{x}\left(\bd{a},\bd{b}\right)$ is defined as
\beq\label{V-function}
    V_x(\bd{a}, \bd{b})=q^{2 Q(\bd{a}, \bd{b})-Q(\bd{a}, \bd{a})-Q(\bd{b}, \bd{b})}
    \left(\frac{q}{x}\right)^{|\bd{a}|-|\bd{b}|} \frac{\left(x^2 ; q^2\right)_{|\bd{a}|-|\bd{b}|}}
    {\prod\limits_{i=1}^{n-1}\left(q^2 ; q^2\right)_{a_i-b_i}}.
\eeq
This function was  introduced in \cite{BS2023} in the formulation of Ising-type model related
to the $U_q(\widehat{sl_n})$ algebra, and the use of it will allow us to write down a number of
relations below in a more compact form.

Then we can express the symmetries, proven in \cite{BM2016} from the properties of the 3D $R$-matrix,
in terms of the symmetric $R$-matrix (\ref{Symmetric R-matrix}) and the function (\ref{V-function}) as follows
\beq\label{First symmerty}
    \left[\bar{R}_{I, J}(x)\right]_{\bd{i}, \bd{j}}^{\bd{i}^{\prime}, \bd{j}^{\prime}}=
    \left[\bar{R}_{J, I}(x)\right]_{\tau \bd{j}, \tau \bd{i}}^{\tau \bd{j}^{\prime}, \tau \bd{i}^{\prime}},
\eeq
\beq\label{Second symmetry}
         \left[\bar{R}_{I, J}(x)\right]_{\bd{i}, \bd{j}}^{\bd{i}^{\prime}, \bd{j}^{\prime}}=
         q^{\left[\bd{i^{\prime}},\bd{j^{\prime}}\right]-\left[\bd{i},\bd{j}\right]}\dfrac{V_{q^{-I}}
         (\bd{i}, 0)V_{q^{-J}}(\bd{j}, 0)}{V_{q^{-I}}(\bd{i}^{\prime}, 0)V_{q^{-J}}(\bd{j}^{\prime}, 0)}
         \left[\bar{R}_{I, J}(x)\right]_{\tau \bd{i}^{\prime}, \tau \bd{j}^{\prime}}^{\tau \bd{i}, \tau \bd{j}},
    \eeq
where the indices are related by the permutation  $\tau$ defined in (\ref{tau}).
Here we introduced an additional operation on indices
\beq\label{convolution}
    \left[\bd{i},\bd{j}\right]=(\bd{i}, \bd{j})-(I-|\bd{i}|)(J-|\bd{j}|).
\eeq
We also have another symmetry based on the transformation $\sigma$ from (\ref{sigma}):
\beq\label{Third symmetry}
    \left[\bar{R}_{I, J}(x)\right]_{\bd{i}, \bd{j}}^{\bd{i}^{\prime}, \bd{j}^{\prime}}=
    x^{(|\bd{i}|+|\sigma\bd{i}|)-(|\bd{i^\prime}|+|\sigma\bd{i^{\prime}}|)}\left[\bar{R}_{I, J}
    (x)\right]_{\sigma\bd{i}, \sigma\bd{j}}^{\sigma\bd{i}^{\prime}, \sigma\bd{j}^{\prime}}.
\eeq

Now let us recall the other basic properties of the stochastic $R$-matrix (\ref{R-matrix}):
\begin{itemize}
    \item Stochastic condition
    \beq\label{Stochastic condition}
       \sum_{\bd{i},\bd{j}} \left[S_{I, J}(x)\right]_{\bd{i}, \bd{j}}^{\bd{i}^{\prime}, \bd{j}^{\prime}}=1.
    \eeq
\end{itemize}
This identity is based on the following summation rule:
\beq\label{summation rule}
    \sum_{\bd{\gamma}}\Phi_q(\bd{\gamma} \mid \bd{\beta} ; \la, \mu)=1,
\eeq
which will be important for us in describing stochastic processes. The proof of this fact can be done by
induction on $n$ and using the $q$-Vandermonde summation formula from  \ref{Basic summation formulas}
(see \cite{KMMO16} for details). As a consequence of the summation formula (\ref{summation rule}),
we directly obtain the stochastic condition (\ref{Stochastic condition}).
\begin{itemize}
    \item Regularity
        \beq\label{Regularity condition for stochastic R-matrix}
        S_{I,I}(1) = P_{12},
    \eeq
\end{itemize}
    where $P$ is a permutation operator, acting on $V^{(n)}_I\otimes V^{(n)}_I$. This property can be easily checked
    from the definition of the $R$-matrix (\ref{R-matrix}).
\begin{itemize}
    \item Unitarity condition
    \beq\label{Unitarity condition for stochastic R-matrix}
        S_{I,J}(x) S_{J,I}\left(x^{-1}\right)=I \otimes I.
    \eeq
\end{itemize}
\begin{itemize}
    \item Crossing unitarity condition
        \beq\label{Crossing unitarity condition for stochastic R-matrix}
        M_1 S_{12}^{t_1}(x) M_1^{-1} S_{21}^{t_1}\left(\left(q^n x\right)^{-1}\right)=g_{I J}^{(n)}(x) I \otimes I,
    \eeq
\end{itemize}
    where $M$ is the diagonal matrix
    \beq\label{M-matrix}
    M_{\bd{i}}^{\bd{j}}=q^{2\sum\limits_{k=1}^{n-1}(n-k)i_k}\delta_{\bd{i}}^{\bd{j}},
    \eeq
    and the scalar function in the right-hand side is
    \beq\label{Crossing function}
    g_{I J}^{(n)}(x)=\dfrac{\left(x^2q^{2-I-J};q^2\right)_I\left(x^2q^{2n-I+J};q^2\right)_I}
    {\left(x^2q^{2-I+J};q^2\right)_I\left(x^2q^{2n-I-J};q^2\right)_I}.
    \eeq
To verify that the scalar factor in the unitarity condition (\ref{Unitarity condition for stochastic R-matrix})
is indeed equal to $1$, one needs to apply the stochastic property (\ref{Stochastic condition}).
As mentioned in the previous section,
\eqref{Crossing unitarity condition for stochastic R-matrix}
follows from the general theory of quantum groups.
Using the formula (\ref{Sum of positive roots}) and the fact that the dual Coxeter number $h^{\vee}=n$
for the Lie algebra $sl_n(\mathbb{C})$, we come to the expression in the left-hand side of
(\ref{Crossing unitarity condition for stochastic R-matrix}).

It remains to show that the function in the right-hand side of the crossing unitarity relation is given by (\ref{Crossing function}).
It is enough to check the relation
\beq\label{Check crossing}
    \sum\limits_{\bd{i^{\prime}}, \bd{j^{\prime}}}q^{2\sum\limits_{k=1}^{n-1}(n-k)(i^{\prime\prime}_k-i^{\prime}_k)}
    \left[S_{I, J}(x)\right]_{\bd{i^{\prime}}, \bd{j}}^{\bd{i}^{\prime\prime}, \bd{j}^{\prime}}\left[S_{J, I}
    \left((q^{n}x)^{-1}\right)\right]_{\bd{j^{\prime}}, \bd{i}}^{\bd{j}^{\prime\prime}, \bd{i}^{\prime}}=
    g_{I J}^{(n)}(x)\delta_{\bd{i}}^{\bd{i^{\prime\prime}}}\delta_{\bd{j}}^{\bd{j^{\prime\prime}}}
\eeq
for any diagonal element, say, for $\bd{i}=\bd{j}=\bd{i^{\prime\prime}}=\bd{j^{\prime\prime}}=0$ ($\bd{i^{\prime}}$ and
$\bd{j^{\prime}}$ are internal summation indices). Note that the $R$-matrix contains an internal summation
(see definition (\ref{R-matrix})). However, we can  remove internal sums by applying the symmetry (\ref{First symmerty})
to both factors in (\ref{Check crossing}), since in this case the $R$-matrix contains only one term. Then one can
rewrite \eqref{Check crossing} using (\ref{phi-function alternative}), and we come to the formula
\beq
\bed
   \sum\limits_{\bd{j^{\prime}}}q^{-2\sum\limits_{k=1}^{n-1}k j^{\prime}_k+n|\bd{j^{\prime}}|}
   \frac{V_{q^{-J}}(\bd{j^{\prime}},0) V_{q^{-I}}(\bd{j^{\prime}},0)}{V_{\frac{q^{-\frac{I+J}{2}}}{x}}
   (\bd{j^{\prime}},0)V_{x q^{-\frac{I+J}{2}+n}}(\bd{j^{\prime}},0)}= g_{I J}^{(n)}(x).
\eed
\eeq
This identity is a consequence of the equation (\ref{first sum of V}), that will be considered in the following sections.
And this ends the proof of the crossing unitarity relation.

In what follows, we will also need the expressions for $L$-operators, which we consider as $n\times n$ matrices acting
in the symmetric tensor representation of $U_q(\widehat{sl_n})$. Namely, we introduce
$(n-1)$-component vectors {$e_0=j_0=(0,\dots,0)$} (vectors of zeros) and $e_\alpha=(0,\dots,0,1,0,\dots,0)$ with $1$
at the $\alpha$-th position ($\alpha=1,\dots, n-1$). Then we construct $L$-operators
acting on $V_1^{(n)}\otimes V_J^{(n)}$ and on $V_J^{(n)}\otimes V_1^{(n)}$, respectively:
\beq\label{L-operators 1,J}
\bed
    &\left[S_{1, J}(x)\right]_{e_\alpha, \bd{j}}^{e_\beta, \bd{l}}=\\& \begin{cases}\delta_{\bd{j},\bd{l}}
    q^{2|e_\alpha|(\sum\limits_{s=1}^{\alpha}j_s-J)} \frac{1-q^{J-1-2j_\alpha-2\delta_{\alpha,0}(J-|\bd{j}|)}
    x^{-2}}{1-q^{-1-J}x^{-2}} & \text { if } \alpha=\beta, \\ -\delta_{\bd{j}+e_\alpha,\bd{l}+e_\beta}
    q^{2(\sum\limits_{s=1}^{\alpha-1}j_s-J-1)+\delta_{\beta,0}(J+1)} x^{-2\delta_{\beta,0}} \frac{1-q^{2+2j_\alpha}}
    {1-q^{-1-J}x^{-2}} & \text { if } \alpha>\beta, \\ -\delta_{\bd{j}+e_\alpha,\bd{l}+e_\beta}
    q^{2\sum\limits_{s=1}^{\alpha-1}j_s-1-J-\delta_{\alpha,0}(J+1-2|\bd{j}|)}x^{-2|e_\alpha|}
    \frac{1-q^{2(1+j_\alpha+\delta_{\alpha,0}(J-|\bd{j}|))}}{1-q^{-1-J}x^{-2}} & \text { if } \alpha<\beta,\end{cases}
\eed
\eeq
\beq\label{L-operators, J,1}
\bed
    &\left[S_{J, 1}(x)\right]_{\bd{j}, e_\alpha}^{\bd{l}, e_\beta}=\\& \begin{cases}\delta_{\bd{j},\bd{l}}
    q^{-2|\bd{j}|+2|e_\alpha|\sum\limits_{s=\alpha}^{n-1}j_s} \frac{1-q^{J-1-2j_\alpha-2\delta_{\alpha,0}
    (J-|\bd{j}|)}x^{-2}}{1-q^{-1-J}x^{-2}} & \text { if } \alpha=\beta, \\ -\delta_{\bd{j}+e_\alpha,\bd{l}+e_\beta}
    q^{-1+J-\delta_{\beta,0}(1+J)-2\sum\limits_{s=1}^{\alpha}j_s} x^{-2|e_\beta|} \frac{1-q^{2+2j_\alpha}}
    {1-q^{-1-J}x^{-2}} & \text { if } \alpha>\beta, \\ -\delta_{\bd{j}+e_\alpha,\bd{l}+e_\beta}
    q^{-2+\delta_{\alpha,0}(1-J)-2|e_\alpha|\sum\limits_{s=1}^{\alpha}j_s}x^{-2\delta_{\alpha,0}}
    \frac{1-q^{2+2j_\alpha+2\delta_{\alpha,0}(J-|\bd{j}|)}}{1-q^{-1-J}x^{-2}} & \text { if } \alpha<\beta.\end{cases}
\eed
\eeq
These relations can be derived directly from the definition of the $R$-matrix (\ref{R-matrix}).

\section{The star-star relation}\label{Section 4}
Using any of the symmetries (\ref{First symmerty}), (\ref{Second symmetry}),
 one can obtain the star-star relation for $U_q(\widehat{sl_n})$,
 written in terms of the functions (\ref{V-function}):
   \beq\label{star-star relation}
        \bed
&\frac{V_{y / y^{\prime}}(\bd{b}, \bd{d})}{V_{y / y^{\prime}}(\bd{a}, \bd{c})} \sum_{\bd{m}=
\max (\bd{b}, \bd{c})}^{\bd{a}} \frac{V_{x / y^{\prime}}(\bd{a}, \bd{m})
V_{y^{\prime} / x^{\prime}}(\bd{m}, \bd{b}) V_{y / x}(\bd{m}, \bd{c})}
{V_{y / x^{\prime}}(\bd{m}, \bd{d})}=\\
& =\frac{V_{x / x^{\prime}}(\bd{a}, \bd{b})}{V_{x / x^{\prime}}(\bd{c}, \bd{d})}
\sum_{\bd{m}=\bd{d}}^{\min (\bd{b}, \bd{c})} \frac{V_{x / y^{\prime}}(\bd{m}, \bd{d})
V_{y^{\prime} / x^{\prime}}(\bd{c}, \bd{m}) V_{y / x}(\bd{b}, \bd{m})}{V_{y / x^{\prime}}(\bd{a}, \bd{m})},
\eed
    \eeq
 which was presented by Bazhanov and Sergeev in \cite{BS2023} as a conjecture. Namely, let us introduce
 the new matrix indices
    \beq
        \bd{i}=\bd{b}-\bd{d}, \quad \bd{j}=\bd{a}-\bd{b}, \quad \bd{i}^{\prime}=\bd{a}-\bd{c},
        \quad \bd{j}^{\prime}=\bd{c}-\bd{d},
    \eeq
    which automatically imply the conservation law $\bd{i}+\bd{j}=\bd{i}^{\prime}+\bd{j}^{\prime}$ coming from
    the Kronecker delta-symbol in the definition of the $R$-matrix (\ref{R-matrix}).

    Now let us note the property of the function $\Phi$ with permuted indices
      \beq\label{Phi-function permuted indices}
    \Phi_q(\tau\bd{\gamma} \mid \tau\bd{\beta} ; \la, \mu)=q^{Q(\bd{\gamma},\bd{\beta})-Q(\bd{\beta},
    \bd{\gamma})}\Phi_q(\bd{\gamma} \mid \bd{\beta} ; \la, \mu)
    \eeq
    and the property of the function $V$
    \beq\label{V-function property 1}
         V_x(\bd{a}, \bd{b})=q^{Q(\bd{a},\bd{b})-Q(\bd{b},\bd{a})}V_x(\bd{a}-\bd{b}, 0),
    \eeq
    which can be easily verified from the corresponding definitions (\ref{phi-function}), (\ref{V-function}).

 If we take (\ref{First symmerty}) or (\ref{Second symmetry}) and use \eqref{phi-function alternative}
 together with \eqref{Phi-function permuted indices} and \eqref{V-function property 1},
    one can see that all pre-factors on both sides of  (\ref{First symmerty}) or (\ref{Second symmetry})
    cancel out, and we come to the star-star relation \eqref{star-star relation}.

\section{Stochastic right boundaries}\label{Right boundaries}\label{Section 5}
In this section we construct special cases of the right integrable boundaries,
which are the upper-triangular solutions to the reflection equation and those that are
generated from the former by applying symmetries of the stochastic $R$-matrix.

The reflection equation for the stochastic $R$-matrix (\ref{R-matrix}) has the form
\begin{equation}\label{Reflection equation for stochastic R-matrix}
    S_{I,J}\left(\frac{x}{y}\right) K_I\left(x\right) S_{J,I}\left(x y\right) K_J\left(y\right)=
    K_J\left(y\right) S_{I,J}\left(x y\right) K_I\left(x\right) S_{J,I}\left(\frac{x}{y}\right).
\end{equation}
First, we show that for a solution $K_J^{(1)}(y)$ of
\eqref{Reflection equation for stochastic R-matrix},
we can construct another solution
\begin{equation}\label{K_sj^sl}
    K_J^{(2)}(y)_{\bd{j}}^{\bd{l}}=\left(\frac{1}{y^2 \mu  q^J}
    \right)^{-|\sigma\bd{j}|}\left(\frac{y^2}{\mu  q^J}\right)^{|\sigma\bd{l}|}
    K_J^{(1)}(y)_{\sigma\bd{j}}^{\sigma\bd{l}},
\end{equation}
where $\mu\in\mathbb{C}$ is an arbitrary parameter. To prove this statement, we will write the reflection equation
(\ref{Reflection equation for stochastic R-matrix}) for $K_J^{(2)}(y)$ in index notation, but with both external and internal
summation indices replaced by $j\rightarrow \sigma^{-1}j$, defined in   (\ref{sigma}),
\begin{equation}\label{trans}
    \begin{aligned}
& \sum_{\substack{\bd{i}^{\prime},  \bd{j}^{\prime}, \\ \bd{i}^{\prime \prime},
\bd{j}^{\prime \prime}}}\left[S_{I, J}\left(\frac{x}{y}\right)\right]_{\sigma^{-1} \bd{i},
\sigma^{-1} \bd{j}}^{\sigma^{-1} \bd{i}^{\prime}, \sigma^{-1} \bd{j}^{\prime}}
\left[K_I^{(2)}(x)\right]_{\sigma^{-1} \bd{i}^{\prime}}^{\sigma^{-1} \bd{i}^{\prime \prime}}
\left[S_{J, I}(x y)\right]_{\sigma^{-1} \bd{j}^{\prime}, \sigma^{-1} \bd{i}^{\prime
 \prime}}^{\sigma^{-1} \bd{j}^{\prime \prime}, \sigma^{-1} \bd{i}^{\prime \prime \prime}}
 \left[K_J^{(2)}(y)\right]_{\sigma^{-1} \bd{j}^{\prime \prime}}^{\sigma^{-1} \bd{j}^{\prime \prime \prime}}= \\
& \sum_{\substack{\bd{i}^{\prime}, \bd{j}^{\prime}, \\ \bd{i}^{\prime \prime},
 \bd{j}^{\prime \prime}}}\left[K_J^{(2)}(y)\right]_{\sigma^{-1} \bd{j}}^{\sigma^{-1}
 \bd{j}^{\prime}}\left[S_{I, J}(x y)\right]_{\sigma^{-1} \bd{i}, \sigma^{-1}
 \bd{j}^{\prime}}^{\sigma^{-1} \bd{i}^{\prime}, \sigma^{-1} \bd{j}^{\prime \prime}}
 \left[K_I^{(2)}(x)\right]_{\sigma^{-1} \bd{i}^{\prime}}^{\sigma^{-1} \bd{i}^{\prime \prime}}
 \left[S_{J, I}\left(\frac{x}{y}\right)\right]_{\sigma^{-1} \bd{j}^{\prime \prime}, \sigma^{-1}
 \bd{i}^{\prime \prime}}^{\sigma^{-1} \bd{j}^{\prime \prime \prime}, \sigma^{-1} \bd{i}^{\prime \prime \prime}}
\end{aligned}
\end{equation}

Now let us note that the $R$-matrix \eqref{R-matrix} possesses another symmetry
\begin{equation}\label{tau-sigma symmetry for S-matrix}
    \left[S_{I, J}\left(x\right)\right]_{\bd{i}, \bd{j}}^{\bd{i}^{\prime}, \bd{j}^{\prime}}=
    \left[S_{J, I}\left(x\right)\right]_{\tau\sigma\bd{j}, \tau\sigma\bd{i}}^{\tau\sigma\bd{j}^{\prime},
    \tau\sigma\bd{i}^{\prime}},
\end{equation}
which follows from the relations (\ref{First symmerty}) and (\ref{Third symmetry}).

Applying \eqref{tau-sigma symmetry for S-matrix} and
(\ref{First symmerty}) to \eqref{trans} and using the fact that
\begin{equation}
    \left[K^{(2)}_J(y)\right]_{\sigma^{-1}\bd{j}}^{\sigma^{-1}\bd{l}}=\left(\frac{1}{y^2 \mu  q^J}
     \right)^{-|\bd{j}|}\left(\frac{y^2}{\mu  q^J}\right)^{|\bd{l}|}\left[K^{(1)}_J(y)\right]_{\bd{j}}^{\bd{l}},
\end{equation}
one can check that all pre-factors on both sides of \eqref{trans} cancel, and we come to the reflection
equation for $K_J^{(1)}(y)$.

Using the symmetry (\ref{K_sj^sl}) $k$ times, $k=1,\ldots,n-1$,
we will obtain
a set of  different solutions to the reflection equation (\ref{Reflection equation for stochastic R-matrix}).
Moreover, we can adjust the free parameter $\mu\in\mathbb{C}$ in \eqref{K_sj^sl} at each step to satisfy
the stochastic condition.

For example, the solutions $K^{(2,3)}(x)$ from  \cite{Crampe14} can be obtained from $K^{(1)}(x)$ by using the
symmetry (\ref{K_sj^sl}) for $n=3$. However, the solution $K^{(4)}$ can not be obtained
in this way.

 Now let us construct the upper-triangular boundary matrices. We will use the same approach as described
 in \cite{Man2019} for the $U_q(\widehat{sl_2})$ case. Namely, we will start from considering the
 reflection equation for $I=J=1$ and show that it admits the solution
\beq\label{Upper-triangular solution, J=1}
    K_1(y)_{\bd{j}}^{\bd{l}}=\Phi_{q^2}\left(\bd{j}|\bd{l};\dfrac{y^2}{ \nu q},\dfrac{1}{y^2 \nu q}\right),
\eeq
which is a generalization for $U_q(\widehat{sl_n})$ with arbitrary $n$. Here $\nu\in\mathbb{C}$ is
a free parameter. Then, using this solution, we will construct a system of linear recurrence equations for
$I=1$ and arbitrary $J.$ And, at the final step, we will show that an upper-triangular stochastic solution
to these equations has the form
\beq\label{Upper-triangular solution}
K_J(y)_{\bd{j}}^{\bd{l}}=\Phi_{q^2}\left(\bd{j}|\bd{l};\dfrac{y^2}{ \nu q^{J}},\dfrac{1}{y^2 \nu q^{J}}\right),
\eeq
which satisfies the stochastic condition
\beq\label{Stochasticity of K-matrix}
    \sum\limits_{\bd{j}}K_J(y)_{\bd{j}}^{\bd{l}}=1,
\eeq
as can be seen from the formula (\ref{summation rule}).

It is convenient to consider the reflection equation (\ref{Reflection equation for stochastic R-matrix})
in a matrix form, that is, to use the expressions for $L$-operators (\ref{L-operators 1,J}) and
(\ref{L-operators, J,1}). In a particular case $J=1$, we come to a simpler formula
\beq\label{L-operators, 1,1}
      \left[S_{1, 1}(x)\right]_{e_\alpha, e_\gamma}^{e_\beta, e_\delta}= \begin{cases}\delta_{\gamma,\delta}
      q^{2Q(e_\gamma,e_\alpha)-2(1-\delta_{\alpha,\gamma})|e_\alpha|} \frac{1-q^{-2\delta_{\alpha,\gamma}}
      x^{-2}}{1-q^{-2}x^{-2}} & \text { if } \alpha=\beta, \\ -\delta_{e_\alpha+e_\gamma,e_\beta+e_\delta}
      q^{-2} x^{-2\delta_{\beta,0}} \frac{1-q^{2}}{1-q^{-2}x^{-2}} & \text { if } \alpha>\beta, \\
      -\delta_{e_\alpha+e_\gamma,e_\beta+e_\delta}q^{-2}x^{-2|e_\alpha|}\frac{1-q^{2}}{1-q^{-2}x^{-2}} &
      \text { if } \alpha<\beta.\end{cases}
\eeq
The last expression allows us to analyze the structure of the reflection equation for $I=J=1$. So we want
to show that the right boundary (\ref{Upper-triangular solution, J=1}) satisfies the relation
\beq\label{Reflection equation, I=J=1}
           \bed
&\sum\limits_{e_{\alpha^{\prime}},e_{\beta^{\prime}},e_{\alpha^{\prime\prime}},e_{\beta^{\prime\prime}}}
\left[S_{1, 1}\left(\dfrac{x}{y}\right)\right]_{e_\alpha, e_\beta}^{e_{\alpha^{\prime}}, e_{\beta^{\prime}}}
\left[K_1\left(x\right)\right]_{e_{\alpha^{\prime}}}^{e_{\alpha^{\prime\prime}}}\left[S_{1, 1}
\left(x y\right)\right]_{e_{\beta^{\prime}},e_{\alpha^{\prime\prime}}}^{e_{\beta^{\prime\prime}},
e_{\alpha^{\prime\prime\prime}}}\left[K_1\left(y\right)\right]_{e_{\beta^{\prime\prime}}}^{e_{\beta^{\prime\prime\prime}}}=\\
&=\sum\limits_{e_{\alpha^{\prime}},e_{\beta^{\prime}},e_{\alpha^{\prime\prime}},e_{\beta^{\prime\prime}}}
\left[K_1\left(y\right)\right]_{e_{\beta}}^{e_{\beta^{\prime}}}\left[S_{1, 1}
\left(x y\right)\right]_{e_{\alpha},e_{\beta^{\prime}}}^{e_{\alpha^{\prime}}, e_{\beta^{\prime\prime}}}
\left[K_1\left(x\right)\right]_{e_{\alpha^{\prime}}}^{e_{\alpha^{\prime\prime}}}\left[S_{1, 1}
\left(\dfrac{x}{y}\right)\right]_{e_{\beta^{\prime\prime}}, e_{\alpha^{\prime\prime}}}^{e_{\beta^{\prime\prime\prime}},
e_{\alpha^{\prime\prime\prime}}},
\eed
\eeq
where the summation goes over internal repeated indices, and the upper-triangular matrix $K_1(x)$ has the form
\beq\label{Upper-triangular solution, in indices}
\left[K_1\left(x\right)\right]_{e_{\alpha}}^{e_{\beta}}=\begin{cases} 1 & \text{ if } \alpha=\beta=0,\\
\frac{q \nu(1-x^4)}{x^2(1-q \nu x^2)} & \text { if } \alpha=0, \beta\ne 0, \\ 0 & \text { if } \alpha\ne 0, \beta=0, \\
\delta_{\alpha,\beta}\frac{x^2-q \nu}{x^2(1-q \nu x^2)} & \text { if } \alpha\ne0, \beta\ne 0.
\end{cases}
\eeq
It is clear from the formulas (\ref{L-operators, 1,1}) and (\ref{Upper-triangular solution, in indices})
that the reflection equation (\ref{Reflection equation, I=J=1}) has the same form for any $n$: indeed,
it only depends on the external indices $e_\alpha$, $e_\beta$, $e_{\alpha^{\prime\prime\prime}}$,
$e_{\beta^{\prime\prime\prime}}$ (more precisely, on the relative order of the numbers $\alpha$,
 $\beta$, $\alpha^{\prime\prime\prime}$, $\beta^{\prime\prime\prime}$), and each factor in the relation
 gives no more than 2 nontrivial terms. This means that it is enough to check the correctness of the equation for,
 say, $n=5$ (to consider all possibilities for index permutations). And, indeed, the explicit calculation shows
 that the equation is valid in this case, which proves (\ref{Upper-triangular solution, J=1}).

Now, based on the solution for $I=1$, we can consider the reflection equation for an arbitrary parameter $J$
\beq\label{Reflection equation, I=1, J}
    S_{1,J}\left(\frac{x}{y}\right) K_1\left(x\right) S_{J,1}\left(x y\right) K_J\left(y\right)=K_J
    \left(y\right) S_{1,J}\left(x y\right) K_1\left(x\right) S_{J,1}\left(\frac{x}{y}\right),
\eeq
where $K_1(x)$ is given by (\ref{Upper-triangular solution, J=1}). Then, substituting (\ref{L-operators 1,J}),
(\ref{L-operators, J,1}) in \eqref{Reflection equation, I=1, J} in matrix form and performing some cumbersome calculations,
we obtain a system of $n^2$ linear equations, that are polynomial in $x$ and $y$. Decoupling with respect to $x$,
we get a set of recurrence relations for $K_J(y)_{\bd{j}}^{\bd{l}}.$ A careful analysis of these equations
shows that there are $n$ linearly independent ones:
\beq\label{Independent equation_1}
\bed
    \nu \left(1-q^{2l_1}\right)\left[K_J(y)\right]_{\bd{j}}^{\bd{l}-e_1}&-q^{-J} y^2 \left(q^{2 j_1}-q^{2 l_1}\right)
    \left[K_J(y)\right]_{\bd{j}}^{\bd{l}}-\\&-\nu  y^4 \left(1-q^{2 j_1+2}\right)\left[K_J(y)\right]_{\bd{j}+e_1}^{\bd{l}}=0,
\eed
\eeq
\beq\label{Independent equation_2}
\bed
    &\left(q^{-2\sum\limits_{k=1}^{i}l_k}-q^{-2\sum\limits_{k=1}^{i}j_k}\right)\left[K_J(y)\right]_{\bd{j}}^{\bd{l}}-
    \sum\limits_{m=1}^{i}q^{-2\sum\limits_{k=1}^{m}l_k}\left(1-q^{2l_m}\right)
    \left[K_J(y)\right]_{\bd{j}}^{\bd{l}-e_m+e_{i+1}}\\&+q^{-2}\sum\limits_{m=1}^{i}q^{-2\sum\limits_{k=1}^{m}j_k}
    \left(1-q^{2+2j_m}\right)\left[K_J(y)\right]_{\bd{j}+e_m-e_{i+1}}^{\bd{l}}=0, \quad\quad i=1,\dots, n-2,
\eed
\eeq
\beq\label{Independent equation_3}
\bed
    \left(q^{-2 j_{n-1}}-q^{-2 l_{n-1}}\right)\left[K_J(y)\right]_{\bd{j}}^{\bd{l}}&-\left(1-q^{-2 l_{n-1}}\right)
    \left[K_J(y)\right]_{\bd{j}}^{\bd{l}-e_{n-1}}+\\&+\left(1-q^{-2-2j_{n-1}}\right)\left[K_J(y)\right]_{\bd{j}+e_{n-1}}^{\bd{l}}=0,
\eed
\eeq
    where we assume that $\left[K_J(y)\right]_{\bd{j}}^{\bd{l}}=0$ unless $j_\alpha\ge 0$, $l_\alpha\ge 0$,
    $|\bd{j}|\le J$ and $|\bd{l}|\le J.$ And in the last step one can check that (\ref{Upper-triangular solution})
    solves these recurrence equations. Thus we constructed the integrable upper-triangular right boundary solution.

Applying (\ref{K_sj^sl}) to (\ref{Upper-triangular solution})
and choosing $\mu=\nu$,  we obtain a stochastic lower-triangular solution of the form
\beq\label{K-matrix, lower-triangular solution}
\bed
    K_J(y)_{\bd{j}}^{\bd{l}}&=\left(\frac{1}{y^2 \nu  q^J} \right)^{-|\sigma\bd{j}|}
    \left(\frac{y^2}{\nu  q^J}\right)^{|\sigma\bd{l}|}\Phi_{q^2}\left(\sigma\bd{j}\mid\sigma\bd{l};
    \frac{y^2}{\nu  q^J},\frac{1}{y^2 \nu  q^J} \right)=\\&=\widehat{\Phi}_{q^2}\left(\sigma\bd{l}-\sigma\bd{j}\mid
    \sigma\bd{l}; y^{-4}, \frac{1}{y^2 \nu  q^J} \right),
\eed
\eeq
where the operation on indices $\sigma$ is defined in (\ref{sigma}), and we introduced a new notation
\beq\label{Phi-hat}
    \widehat{\Phi}_{q}\left(\bd{j}\mid \bd{l}; x, y \right)=q^{Q(\bd{j},  \bd{l})-Q(\bd{l},
    \bd{j})}\Phi_{q}\left(\bd{j}\mid \bd{l}; x, y \right).
\eeq
In the second line in (\ref{K-matrix, lower-triangular solution}) we applied the relation.
\beq\label{Relation for phi-function}
    \Phi_q(\bd{m}-\bd{j} \mid \bd{m}, \mu / \la, \mu)=\Phi_q(\bd{j} \mid \bd{m}, \la, \mu)
    q^{Q(\bd{j},\bd{m})-Q(\bd{m},\bd{j})} \mu^{-|\bd{j}|} \la^{|\bd{m}|} .
\eeq

The stochastic condition (\ref{Stochasticity of K-matrix}) for the lower-triangular solution follows from the identity
\beq
\bed
    &\sum\limits_{\bd{j}}x^{|\bd{l}|}y^{-|\bd{j}|}\Phi_{q}\left(\bd{j}\mid\bd{l}; x,y \right)=
    \sum\limits_{\bd{j}}q^{Q(\bd{j},\bd{l})-Q(\bd{l},\bd{j})}\Phi_{q}\left(\bd{j}\mid\bd{l}; \frac{y}{x}, y \right)=\\
    &=\sum\limits_{\bd{j}}\Phi_{q}\left(\tau\bd{j}\mid\tau\bd{l}; \frac{y}{x}, y \right)=1,
 \eed
\eeq
where we used the relations (\ref{Relation for phi-function}), (\ref{Phi-function permuted indices}) and
the summation rule (\ref{summation rule}).

It is interesting to compare the obtained solutions (\ref{Upper-triangular solution}) and (\ref{K-matrix, lower-triangular solution}) 
with already known results.{In \cite{he2022shift} the author introduced a solution to the reflection equation related
 to the stochastic 
 $R$-matrix 
\beq 
 \left[\widetilde{S}_{I, J}(x)\right]_{\boldsymbol{i}, \boldsymbol{j}}^{\boldsymbol{i}^{\prime},\boldsymbol{j}^{\prime}}=
 \left[S_{I, J}(x^{-1}q^{\frac{I-J}{2}})\right]_{\sigma^{-1}\boldsymbol{i}, 
 \sigma^{-1}\boldsymbol{j}}^{\sigma^{-1}\boldsymbol{i}^{\prime},\sigma^{-1}\boldsymbol{j}^{\prime}},
 \eeq
 derived in \cite{Borodin_2022} through the fusion procedure. 
 The solution from \cite{he2022shift} has a similar structure to 
 \eqref{Upper-triangular solution}
 and also includes the function $\Phi$ as a building block.}

In \cite{Ragoucy2016} a full classification of the right boundaries for the case $J=1$ was given. It is based on the classification
with four particles
of
two special types (four positive integers):  slow ($s_1$, $s_2$) and fast ($f_1$, $f_2$) ones, which satisfy the constraints
    \beq\label{Special particles}
        1 \leq s_1 \leq s_2<f_2 \leq f_1 \leq n \quad \text { and } \quad f_1-f_2=s_2-s_1.
    \eeq
    Note that the authors of \cite{Ragoucy2016} use different index notations, starting from $1$ rather than $0$. Each solution depends on
    two real parameters $\alpha$, $\gamma$ with the choice of two boundary matrices $B^0\left(\alpha, \gamma|s_1,s_2,f_2,f_1 \right)$ and
    $B\left(\alpha, \gamma|s_1,s_2,f_2,f_1 \right).$ Then one can check that the solution (\ref{Upper-triangular solution})
    for $J=1$ corresponds to the case of the matrix $B\left(\alpha, \gamma|s_1,s_2,f_2,f_1 \right)$ with the
     slow $s_1=s_2=1$ and fast $f_1=f_2=n$ species
   and the parameters
    \beq
        \alpha=0, \quad \gamma=\dfrac{1-q}{1-\nu q},
    \eeq
    and the lower-triangular solution (\ref{K-matrix, lower-triangular solution}) corresponds to  the parameters
    \beq
        \alpha=\dfrac{1-q}{1-\nu q}, \quad \gamma=0.
    \eeq

    To illustrate the structure of the right boundary matrices (\ref{Upper-triangular solution}),
    (\ref{K-matrix, lower-triangular solution}), we listed them in \ref{Some explicit matrices}
    for a particular case $n=3$ and $J=2$.

\section{Stochastic left boundaries}\label{Section 6}
Now we will construct the left upper- and lower-triangular boundaries using the results from the previous sections.
The starting point for this calculation is the dual reflection equation
\beq\label{Dual reflection equation, stochastic R-matrix}
\wt{K}_J\left(y\right)\wt S_{J,I}(x y) \wt{K}_I\left(x\right) S_{J,I}\left(\frac{y}{x}\right)=
S_{I,J}\left(\frac{y}{x}\right) \wt{K}_I\left(x\right)\wt S_{I,J}(x y) \wt{K}_J\left(y\right),
\eeq
where as usual
\beq
    \wt S_{I,J}\left(x\right)=\left(\left(S_{I,J}\left(x\right)^{t_2}\right)^{-1}\right)^{t_2}.\label{left1}
\eeq
First, let us notice that the matrix $M$ (\ref{M-matrix}) satisfies the compatibility condition (\ref{YBE6}):
\beq
    \left[M \otimes M, S_{I,J}(x)\right]=0.\label{left2}
\eeq
As discussed in Section \ref{Section 2}, we can find the solution to
\eqref{Dual reflection equation, stochastic R-matrix} in terms of $K_I(x)$ from (\ref{Upper-triangular solution})
\beq\label{YBE9a}
     \wt{K}_I(x)=M^{-1}K_I\left(\dfrac{1}{q^{n/2}x}\right).
\eeq
However, we are also interested in solutions $\bar{K}_I(x)$ of  the reflection equation (\ref{YBE10}) which enter
the definition of the hamiltonian rates, see \eqref{YBE16}. The corresponding reflection equation
takes the form:
\beq \label{Reflection equation_2, stochastic R-matrix }
S_{J, I}\left(\frac{y}{x}\right) \bar{K}_I\left(x\right) S_{I, J}\left(\frac{1}{x y}\right)
\bar{K}_J\left(y\right)=\bar{K}_J\left(y\right) S_{J, I}\left(\frac{1}{x y}\right) \bar{K}_I\left(x\right) S_{I, J}\left(\frac{y}{x}\right).
\eeq
First, let us discuss the symmetries of this equation.
If $\bar{K}_J^{(1)}(y)$ is a solution of \eqref{Reflection equation_2, stochastic R-matrix }, then
\beq\label{left4}
   \bar{K}_J^{(2)}(y)_{\bd{j}}^{\bd{l}}=\left(\frac{y^2}{\mu  q^J}
    \right)^{|\sigma\bd{j}|}\left(\frac{1}{y^2 \mu  q^J}\right)^{-|\sigma\bd{l}|}
    \bar{K}_J^{(1)}(y)_{\sigma\bd{j}}^{\sigma\bd{l}}
\eeq
is also a solution. This can be shown by
using the symmetry \eqref{tau-sigma symmetry for S-matrix} of the $R$-matrix
and performing a calculation similar to \eqref{trans}.

Proceeding in a similar way, one can show that if $K_J(y)$ solves \eqref{Reflection equation for stochastic R-matrix}, then
\begin{equation}
    \bar{K}_J(y)_{\boldsymbol{j}}^{\boldsymbol{l}}= K_J(y^{-1})_{\tau\sigma\boldsymbol{j}}^{\tau\sigma\boldsymbol{l}}\label{left3}
\end{equation}
solves \eqref{Reflection equation_2, stochastic R-matrix }. Let us emphasize that the formula \eqref{left3} is a consequence
of the symmetry \eqref{tau-sigma symmetry for S-matrix}.

However, we also have a general construction for the matrix $\bar{K}(x)$ given by \eqref{YBE12} and \eqref{YBE9}.
In the rest of this section we will show that it gives the same result \eqref{left3} up to the transformation \eqref{left4}.

Combining \eqref{YBE12} with \eqref{YBE9a}, we obtain
\beq\label{Inverse map, stochastic R-matrix}
\bed
    \bar{K}_I(x)=\op{tr}_{V_J}\left(M_{J}^{-1}K_J\left(\dfrac{x}{q^{\frac{n}{2}}}\right)S_{J,I}\left(\dfrac{1}{x^2}\right) P_{I,J}\right),
\eed
\eeq
where we assume that $I=J$ to correctly define the permutation operator acting on two isomorphic spaces.

We start our analysis with the upper-triangular solutions \eqref{Upper-triangular solution}. Substituting
(\ref{M-matrix}) in (\ref{Inverse map, stochastic R-matrix})
and replacing an arbitrary parameter $\nu\rightarrow\dfrac{1}{\nu q^n}$, we obtain the following formula:
\beq\label{index notation}
\bed    \bar{K}_J(x)_{\bd{j}}^{\bd{j}^{\prime}}=\sum\limits_{\bd{i},\bd{i^{\prime}}}q^{-2\sum\limits_{k=1}^{n-1}(n-k)i_k}\left.K_J
\left(\dfrac{x}{q^{n/2}}\right)_{\bd{i}}^{\bd{i}^{\prime}}\right|_{\nu\rightarrow\frac{1}{\nu q^n}}\left[S_{J, J}
\left(\dfrac{1}{x^2}\right)\right]_{\bd{i}^{\prime}, \bd{j}}^{\bd{j}^{\prime}, \bd{i}}.
\eed
\eeq
We claim that \eqref{index notation} can be transformed to the following form
\beq \label{K-matr_upper}
   \bed
\bar{K}_J(y)_{\bd{j}}^{\bd{l}}&=\lambda_J^{(n)}(x)\left(\frac{y^2}{\nu q^J} \right)^{-|\boldsymbol{j}|}
\left(\frac{1}{y^2 \nu  q^J}\right)^{|\boldsymbol{l}|}q^{2\left(Q(\boldsymbol{j},\boldsymbol{l})-
Q(\boldsymbol{l},\boldsymbol{j})\right)}\Phi_{q^2}\left(\boldsymbol{j}|\boldsymbol{l};\dfrac{1}{y^2 \nu q^{J}},
\dfrac{y^2}{\nu q^{J}}\right)=\\&=\la_J^{(n)}(x)\Phi_{q^2}\left(\bd{l}-\bd{j}|\bd{l};y^4,\dfrac{y^2}{\nu q^{J}} \right),
\eed
\eeq
where in the last line we used (\ref{Relation for phi-function}) and
introduced the function
\beq\label{lambda function}
    \la_J^{(n)}(x)=\dfrac{\left(x^{-4}q^{2J+2};q^2\right)_{n-1} \left(\nu x^{-2}q^{2-J};q^2\right)_{n-1}}{\left(x^{-4}q^2;q^2\right)_{n-1}
   \left(\nu x^{-2}q^{J+2};q^2\right)_{n-1}}.
\eeq
Combining the first line in \eqref{K-matr_upper} with the symmetry transformation \eqref{left4}, we see that
the formula \eqref{K-matr_upper} is equivalent to \eqref{left3} up to a scalar factor \eqref{lambda function}.

To prove the equivalence of \eqref{index notation} and \eqref{K-matr_upper}, we want to apply certain
 summation formulas. A simple analysis shows that, by using a symmetry transformation
\beq\label{symmetry for S-matrix}
    \left[S_{J, J}\left(x\right)\right]_{\bd{i}, \bd{j}}^{\bd{i}^{\prime}, \bd{j}^{\prime}}=
    x^{2\left(|\bd{i}^{\prime}|-|\bd{i}|\right)}q^{2J\left(|\bd{j}^{\prime}|-|\bd{i}|\right)}\left[S_{J, J}
    \left(x\right)\right]_{\tau\bd{j}, \tau\bd{i}}^{\tau\bd{j}^{\prime}, \tau\bd{i}^{\prime}}
\eeq
in \eqref{index notation},
we obtain  the terminating series, truncated in a natural way
by $J$.

Rewriting all functions $\Phi$ in terms of $V$'s
and canceling out the pre-factors, we obtain that the equality  of \eqref{index notation} and \eqref{K-matr_upper}
is equivalent to the proof of the relation
\beq\label{two sum}
\bed
    &\la_J^{(n)}(x)\frac{ V_{\frac{q^{-J/2}}{\nu^{1/2}
   x}}(\bd{b},\bd{a})V_{x^2}(\bd{c},\bd{b})}{V_{\frac{x q^{-J/2}}{\nu^{1/2}
   }}(\bd{c},\bd{a})}=\frac{V_{\frac{q^n}{x^2}}(\bd{c},\bd{b})}{V_{q^{-J}}(\bd{c}
   ,\bd{a})}\sum\limits_{\bd{m}}V_{x^2}(\bd{b},\bd{m}) V_{x^2}(\bd{c},\bd{m})
   V_{\frac{q^{-J}}{x^2}}(\bd{m},\bd{a})\times\\ &\times\sum\limits_{\bd{d}}q^{-2
   \sum\limits_{k=1}^{n-1}k(d_k-c_k)+n(|\bd{d}|-|\bd{c}|)}\frac{V_{q^{-J}}(\bd{d},\bd{b}) V_{\nu^{1/2}  x
   q^{-J/2}}(\bd{d},\bd{c})}{V_{\frac{\nu^{1/2}
   q^{-\frac{J}{2}+n}}{x}}(\bd{d},\bd{b})V_{x^2 q^{-J}}(\bd{d},\bd{m})}.
\eed
\eeq
Here we introduced matrix indices
\beq
        \bd{i}=\bd{d}-\bd{c}, \quad \bd{j}=\bd{b}-\bd{a}, \quad \bd{i}^{\prime}=\bd{d}-\bd{b}, \quad \bd{j}^{\prime}=\bd{c}-\bd{a}.
    \eeq
In these notations,
all indices satisfy the constraints
\beq
       0 \le a_i\le m_i \le b_i \le c_i \le d_i, \quad |\bd{d}|\le J.
    \eeq
Two summations in (\ref{two sum}) can be done by using the summation formulas
\beq
\bed\label{first sum of V}
   \sum\limits_{\bd{d}}q^{-2\sum\limits_{k=1}^{n-1}k(d_k-c_k)+n(|\bd{d}|-|\bd{c}|)}&\frac{V_{q^{-J}}
   (\bd{d},\bd{b}) V_{y}(\bd{d},\bd{c})}{V_{z}(\bd{d},\bd{b})V_{\frac{y}{z} q^{n-J}}(\bd{d},\bd{m})}=\\
   &= \mu_J^{(n)}(y,z)\frac{V_{q^{-J}}(\bd{c},\bd{b}) V_{\frac{q^{n}}{z }}(\bd{b},\bd{m})}{V_{\frac{y q^n}{z}}
   (\bd{b},\bd{m}) V_{\frac{z}{y}}(\bd{c},\bd{b}) V_{\frac{q^{n-J}}{z}}(\bd{c},\bd{m})},
\eed
\eeq
where
\beq\label{mu function}
     \mu_J^{(n)}(y,z)=\frac{\left(q^{2-2n} z^2;q^2\right)_{n-1} \left(\frac{q^{2J-2n+2}
   z^2}{y^2};q^2\right)_{n-1}}{\left(q^{2J-2n+2}z^2;q^2\right)_{n-1}
   \left(\frac{q^{2-2 n} z^2}{y^2};q^2\right)_{n-1}},
\eeq
and
\beq\label{second sum of V}
   \dfrac{V_{y / y^{\prime}}(\bd{c},\bd{b})}{V_{y / y^{\prime}}(\bd{c}
   ,\bd{a})}\sum\limits_{\bd{m}}\dfrac{V_{x / y^{\prime}}(\bd{m},\bd{a})
   V_{y^{\prime} / x^{\prime}}(\bd{b},\bd{m})V_{y / x}(\bd{c},\bd{m})}{V_{y / x^{\prime}}
   (\bd{c},\bd{m})}=\frac{ V_{x / x^{\prime}}(\bd{b},\bd{a})
   V_{y / x}(\bd{c},\bd{b})}{V_{y / x^{\prime}}(\bd{c},\bd{a})}.
\eeq
The relation \eqref{first sum of V} is proven in \ref{AppC}. The relation \eqref{second sum of V} easily follows
from the star-star relation \eqref{star-star relation}. Indeed, let us substitute $\bd{b}=\bd{a}$
in \eqref{star-star relation}.
Then the sum in the LHS disappears, since $\bd{a}\geq \bd{m}$ and $\bd{m}\geq\bd{b}=\bd{a}$ and
\eqref{star-star relation} reduces to  \eqref{second sum of V}.

Since the solution
to the reflection equation is defined up to an arbitrary factor,
we divide \eqref{K-matr_upper} by $\la_J^{(n)}(x)$ and obtain for the integrable left upper-triangular boundary
\beq\label{Left boundary}
    \bar{K}_J(y)_{\bd{j}}^{\bd{l}}=\Phi_{q^2}\left(\bd{l}-\bd{j}|\bd{l};y^4,\dfrac{y^2}{\nu q^{J}} \right),
\eeq
which satisfies the stochastic property due to  (\ref{summation rule}).

Now let us construct the stochastic lower-triangular left boundary solutions.
They are obtained by applying the symmetry \eqref{left4} with $\mu=\nu$ to the first line of \eqref{K-matr_upper}
and removing the scalar factor $\la_J^{(n)}(x)$. As a result, we have
\beq\label{Left boundary, lower-triangular solution}
    \bar{K}_J(y)_{\bd{j}}^{\bd{l}}=\widehat{\Phi}_{q^2}\left(\sigma\bd{j}\mid \sigma\bd{l};
    \dfrac{1}{y^2 \nu q^{J}},\dfrac{y^2}{\nu q^{J}} \right),
\eeq
where  $\widehat{\Phi}_{q}$ is defined in (\ref{Phi-hat}).

In \ref{Some explicit matrices}
 we listed left boundary matrices (\ref{Left boundary}),
 (\ref{Left boundary, lower-triangular solution}) for a particular case $n=3$ and $J=2$.

\section{Another model}\label{Section 7}
In this section we will not impose the requirement $I, J\in\mathbb{N}$, i.e the $R$-matrix (\ref{R-matrix})
acts in the tensor product of infinite dimensional modules with basis vectors having the form
\beq
      |\bd{i}\rangle=|i_1,i_2,\dots,i_{n-1}\rangle, \quad i_k\in\mathbb{Z}_{+}.
\eeq
The $R$-matrix (\ref{R-matrix}) degenerates when we choose the spectral parameter as $x=q^{(J-I)/2}$ \cite{BM2016,KMMO16}:
\beq
     \left[S_{I, J}\left(q^{(J-I) / 2}\right)\right]_{\bd{i}, \bd{j}}^{\bd{i}^{\prime}, \bd{j}^{\prime}}=
     \delta_{\bd{i}+\bd{j}, \bd{i}^{\prime}+\bd{j}^{\prime}} \Phi_{q^2}\left(\bd{i} \mid \bd{j}^{\prime} ; q^{-2I}, q^{-2J} \right),
\eeq
which suggests to define a new stochastic non-difference type $R$-matrix
\beq\label{R-matrix, non-difference type}
    \left[S_{12}(x, y)\right]_{\bd{i}, \bd{j}}^{\bd{i}^{\prime}, \bd{j}^{\prime}}=\delta_{\bd{i}+\bd{j},
    \bd{i}^{\prime}+\bd{j}^{\prime}} \Phi_{q^2}\left(\bd{i} \mid \bd{j}^{\prime} ; x, y \right).
\eeq
Now we consider $x, y\in \mathbb{C}$ as spectral parameters and the Yang-Baxter equation takes the form
\beq\label{Yang-Baxter equation, non-difference type}
    S_{12}\left(x, y\right) S_{13}\left(x, z\right) S_{23}\left(y, z\right)=S_{23}\left(y, z\right)
    S_{13}\left(x, z\right) S_{12}\left(x, y\right),
\eeq
which directly follows from ($\ref{Yang-Baxter for stochastic R-matrix}$).
It is easy to see that the internal sums in \eqref{Yang-Baxter equation, non-difference type}
contain only a finite number of terms, so there is no problem with {convergence}.

 For a given solution to the Yang-Baxter equation (\ref{R-matrix, non-difference type}), we define
 the reflection equation of a non-difference type
\beq\label{Reflection equation, non-difference type}
    S_{12}(x, y) K_1(x, \bar{x}) S_{21}(y, \bar{x}) K_2(y, \bar{y})=K_2(y, \bar{y})
    S_{12}(x, \bar{y}) K_1(x, \bar{x}) S_{21}(\bar{y}, \bar{x}) .
\eeq
 In \cite{Man2019} it was suggested that this equation
 admits an upper-triangular stochastic solution of the form
 \beq\label{K-matrix, non-difference type}
     K(x, \bar{x})_{\bd{j}}^{\bd{l}}=\Phi_{q^2}({\bd{j}} \mid {\bd{l}} ; z^2 x, z^2 \bar{x}),
 \eeq
 where $z\in\mathbb{C}$ is an arbitrary parameter
 and there is no constraint between spectral parameters $x$ and $\bar{x}$
  (see the equation (5.10) from \cite{Man2019}).

 Here we give the proof of this fact suggested in \cite{Serg22}, which is based on the star-star relation
 (\ref{star-star relation}) and an orthogonality property:
 \beq\label{Orthogonality}
\sum\limits_{\bd{m}}x^{|\bd{m}|-|\bd{b}|}y^{|\bd{m}|-|\bd{a}|}V_{x}\left(\bd{a},\bd{m}\right)
V_{y}\left(\bd{m},\bd{b}\right)=V_{x y}\left(\bd{a}, \bd{b}\right).
 \eeq
 The last equality directly follows from the relation
 (\ref{summation formula}) and the $q$-Vandermonde summation formula (\ref{q-Vandermonde}). We start the proof
 by writing the reflection equation (\ref{Reflection equation, non-difference type}) in index notation
 \beq\label{Reflection equation,non-difference type, index}
           \bed
&\sum\limits_{\bd{i^{\prime}},\bd{j^{\prime}},\bd{i^{\prime\prime}},\bd{j^{\prime\prime}}}
\left[S_{12}\left(x^2, y^2\right)\right]_{\bd{i}, \bd{j}}^{\bd{i^{\prime}}, \bd{j^{\prime}}}
\left[K_1\left(x^2,\bar{x}^2\right)\right]_{\bd{i^{\prime}}}^{\bd{i^{\prime\prime}}}\left[S_{21}
\left(y^2, \bar{x}^2\right)\right]_{\bd{j^{\prime}},\bd{i^{\prime\prime}}}^{\bd{j^{\prime\prime}},
\bd{i^{\prime\prime\prime}}}\left[K_2\left(y^2, \bar{y}^2\right)\right]_{\bd{j^{\prime\prime}}}^{\bd{j^{\prime\prime\prime}}}=\\
&=\sum\limits_{\bd{i^{\prime}},\bd{j^{\prime}},\bd{i^{\prime\prime}},\bd{j^{\prime\prime}}}
\left[K_2\left(y^2, \bar{y}^2\right)\right]_{\bd{j}}^{\bd{j^{\prime}}}\left[S_{12}
\left(x^2, \bar{y}^2\right)\right]_{\bd{i}, \bd{j^{\prime}}}^{\bd{i^{\prime}},
\bd{j^{\prime\prime}}}\left[K_1\left(x^2,\bar{x}^2\right)\right]_{\bd{i^{\prime}}}^{\bd{i^{\prime\prime}}}
\left[S_{21}\left(\bar{y}^2, \bar{x}^2\right)\right]_{\bd{j^{\prime\prime}},
\bd{i^{\prime\prime}}}^{\bd{j^{\prime\prime\prime}}, \bd{i^{\prime\prime\prime}}},
\eed
\eeq
where, as usual, the summation goes over repeated indices. Then we introduce new external indices
   \beq
        \bd{i}=\bd{b}-\bd{a}, \quad \bd{j}=\bd{c}-\bd{b}, \quad \bd{i^{\prime\prime\prime}}=\bd{e}-\bd{a},
        \quad \bd{j^{\prime\prime\prime}}=\bd{f}-\bd{e},
    \eeq
summation indices for the left-hand side
  \beq
        \bd{i^{\prime}}=\bd{c}-\bd{d}_1, \quad \bd{j^{\prime}}=\bd{d}_1-\bd{a}, \quad \bd{i^{\prime\prime}}=
        \bd{d}_2-\bd{d}_1, \quad \bd{j^{\prime\prime}}=\bd{d}_2-\bd{e}
    \eeq
and for the right-hand side
 \beq
        \bd{i^{\prime}}=\bd{d}_2-\bd{d}_1, \quad \bd{j^{\prime}}=\bd{d}_2-\bd{b}, \quad \bd{i^{\prime\prime}}=
        \bd{f}-\bd{d}_1, \quad \bd{j^{\prime\prime}}=\bd{d}_1-\bd{a}.
    \eeq
After canceling out similar factor on both sides,
the reflection equation in terms of the functions $V_x\left(\bd{a}, \bd{b}\right)$ takes the form
\beq\label{Proof of reflection equation}
\bed
    &\sum\limits_{\bd{d}_2}\frac{\bar{x}^{|\bd{c}|} \bar{y}^{\left|\bd{d}_2\right| }}{x^{\left| \bd{d}_2\right| }
    y^{|\bd{f}|}} \frac{V_{y z}\left(\bd{d}_2,\bd{e}\right) V_{\frac{\bar{x}}{x}}\left(\bd{d}_2,\bd{c}\right)
    V_{\frac{\bar{y}}{y}}\left(\bd{f},\bd{d}_2\right)}{V_{z \bar{y}}(\bd{f},\bd{e})}\sum\limits_{\bd{d}_1}
    \frac{V_{\frac{y}{x}}\left(\bd{d}_1,\bd{b}\right) V_{x z}\left(\bd{c},\bd{d}_1\right) V_{\frac{\bar{x}}{y}}
    \left(\bd{e},\bd{d}_1\right)}{V_{z \bar{x}}\left(\bd{d}_2,\bd{d}_1\right)}=\\
    =&\sum\limits_{\bd{d}_2}\frac{\bar{x}^{\left|\bd{d}_2\right| } \bar{y}^{|\bd{c}| }}{x^{|\bd{f}| }
    y^{\left|\bd{d}_2\right|}}\frac{V_{y z}(\bd{c},\bd{b}) V_{\frac{\bar{x}}{x}}\left(\bd{f},\bd{d}_2\right)
    V_{\frac{\bar{y}}{y}}\left(\bd{d}_2,\bd{c}\right)}{V_{z \bar{y}}\left(\bd{d}_2,\bd{b}\right)}
    \sum\limits_{\bd{d}_1}\frac{V_{\frac{\bar{y}}{x}}\left(\bd{d}_1,\bd{b}\right) V_{x z}\left(\bd{d}_2,\bd{d}_1\right)
    V_{\frac{\bar{x}}{\bar{y}}}\left(\bd{e},\bd{d}_1\right)}{V_{z \bar{x}}\left(\bd{f},\bd{d}_1\right)}.
\eed
\eeq
Then, sequentially applying the star-star relation (\ref{star-star relation}) centered at $\bd{d}_1$ and
orthogonality formula ($\ref{Orthogonality}$) with respect to $\bd{d}_2$ to both sides of the relation
(\ref{Proof of reflection equation}), we come to the conclusion that the left and right sides  coincide:
\beq
    \op{LHS}=\op{RHS}=\frac{V_{\frac{\bar{x}}{x}}(\bd{e},\bd{b}) V_{y z}(\bd{c},\bd{b})}{V_{z \bar{y}}
    (\bd{f},\bd{e})}\sum\limits_{\bd{d}_1}\left(\frac{\bar{y}}{y}\right)^{\left|\bd{d}_1\right|}
    \frac{V_{x z}\left(\bd{d}_1,\bd{e}\right) V_{\frac{\bar{x}}{y}}\left(\bd{d}_1,\bd{c}\right)
    V_{\frac{\bar{y}}{x}}\left(\bd{f},\bd{d}_1\right)}{V_{z \bar{x}}\left(\bd{d}_1,\bd{b}\right)},
\eeq
which ends the proof.

Note that $S_{12}^{t_2}\left(x, y\right)$ is not invertible, and we cannot define $\wt S_{12}(x, y)$ and
therefore the dual reflection equation. However, the inverse $R$-matrix $S_{12}^{-1}\left(x, y\right)$ is well defined
and has the form
\beq\label{Inverse R-matrix}
    S_{12}^{-1}\left(x, y\right)=S_{21}\left(y, x\right),
\eeq
which is a consequence of the orthogonality property (\ref{Orthogonality}). Therefore we can try to define
an analogue of the reflection equation (\ref{YBE10}) based on the inverse $R$-matrix (\ref{Inverse R-matrix}):
\beq\label{Reflection equation_2, non-difference type}
    S_{21}\left(y, x\right) \bar{K}_1\left(x, \bar{x} \right) S_{12}\left(\bar{x}, y\right) \bar{K}_2
    \left(y, \bar{y} \right)=\bar{K}_2\left(y, \bar{y}\right) S_{21}\left(\bar{y}, x\right) \bar{K}_1
    \left(x, \bar{x} \right) S_{12}\left(\bar{x}, \bar{y} \right).
\eeq
From (\ref{Orthogonality}) one can establish that this equation admits a stochastic solution of the form
\beq
    \bar{K}\left(x, \bar{x} \right)=\Phi_{q^2}\left({\bd{l}}-{\bd{j}} \mid {\bd{l}} ; \dfrac{\bar{x}}{x}, \bar{x}\right),
\eeq
which, however, does not include an additional free parameter as in (\ref{K-matrix, non-difference type}).

\section{Conclusion}\label{Section 8}
In this paper we constructed triangular solutions for the reflection equation corresponding to symmetric
tensor representations of $U_q(\widehat{sl_n})$.
It is quite remarkable that all constructed solutions  are built from the same function $\Phi_q(\bd{\gamma} \mid \bd{\beta} ; \la, \mu)$
defined in \eqref{phi-function}.
As a result, both reflection and dual reflection  equations \eqref{Reflection equation for stochastic R-matrix},
\eqref{Dual reflection equation, stochastic R-matrix}  reduce to multiple sum identities
among the $\Phi$-functions (or $V$-functions \eqref{V-function}). It would be natural to expect that
\eqref{Reflection equation for stochastic R-matrix},
\eqref{Dual reflection equation, stochastic R-matrix} can be proven by a repeated application of the star-star relation
\eqref{star-star relation}. However,
we failed to find a correct sequence of the star-star relations which would prove these equations.
This is the reason why we introduced the $L$-operators \eqref{L-operators 1,J} and \eqref{L-operators, J,1} to prove that
the solution \eqref{Upper-triangular solution} solves the reflection equation \eqref{Reflection equation, I=1, J}.
We are not aware of any other nontrivial identities among $V$-functions
except \eqref{star-star relation} and believe that such an algebraic proof  may
still exist.

We expect that general stochastic solutions for the $U_q(\widehat{sl_n})$ algebra
should have one extra parameter as demonstrated in \cite{Man2019} for $U_q(\widehat{sl_2})$ and in  \cite{Ragoucy2016} for $U_q(\widehat{sl_n})$ with $J=1$.
 Since our triangular solutions are factorised,
they may serve as building blocks for general solutions similar to the expression of the stochastic $R$-matrix
\eqref{R-matrix}. However, we have not yet found  such a general representation and aim to do that in the future.

\section{Acknowledgements}

We would like to thank V. Bazhanov and S. Sergeev for careful reading of the manuscript and valuable comments
at different stages of this project.

\appendix

\section{Solution to the dual reflection equation}\label{AppA}
In this Appendix we show that \eqref{YBE11} solves the dual reflection equation \eqref{YBE8} and
prove the inverse mapping \eqref{YBE12},
see also \cite{Vlaar2015} for similar calculations.

First, we notice two different forms of the Yang-Baxter equation
\beq\label{almost Y-B}
  R_{12}(x)\wt{R}_{23}(y) \wt{R}_{13}(xy)=\wt{R}_{13}(xy) \wt{R}_{23}(y)R_{12}(x)
\eeq
\beq\label{almost Y-B2}
  \wt R_{12}(x)\wt{R}_{13}(xy) {R}_{23}(y)={R}_{23}(y) \wt{R}_{13}(xy)\wt R_{12}(x)
\eeq
which can be proved using \labelcref{YBE1,R-tilde,YBE3}.

Now we show that (\ref{YBE11}) solves the dual reflection equation (\ref{YBE8}). First, let us
calculate the product of the first three factors
in the LHS of \eqref{YBE8}
\begin{align}
&\wt{K}_2\left(y\right)\wt R_{21}(x y) \wt{K}_1\left(x\right)=\nonumber\\
&\op{tr}_0\left[\bar{K}_{0}\left(\frac{1}{y}\right)\wt R_{02}(y^2)P_{02} \right]
\wt R_{21}(x y)\op{tr}_{0'}\left[\bar{K}_{0'}\left(\frac{1}{x}\right)\wt R_{0'1}(x^2)P_{0'1} \right]\label{A1}
\end{align}
Using the property $\tr_i A^{t_i}B^{t_i}=\tr_i A B$ and the definition \eqref{R-tilde} , we transform \eqref{A1} to
\bea
&\op{tr}_0\left[\left(\bar{K}_{0}\left(\frac{1}{y}\right)P_{02}\right)^{t_0}
\left(\wt R_{20}(y^2)\right)^{t_0} \right]\wt R_{21}(x y)\op{tr}_{0'}
\left[\bar{K}_{0'}\left(\frac{1}{x}\right)\wt R_{0'1}(x^2)P_{0'1} \right]=\nonumber\\
&\op{tr}\left[\bar{K}_{0'}\left(\frac{1}{x}\right) \left(\bar{K}_{0}\left(\frac{1}{y}\right)P_{02}\right)^{t_0}
P_{0'1}\left(\wt R_{20}(y^2)\right)^{t_0}\wt R_{20'}(x y)\wt R_{10'}(x^2) \right]=\nonumber\\
&\tr\left[\bar{K}_{0'}\left(\frac{1}{x}\right) \left(\bar{K}_{0}\left(\frac{1}{y}\right)P_{02}\right)^{t_0}
R_{0'0}^{t_0}(x y)P_{0'1}\wt R_{10}^{t_0}(x y)\wt R_{20}^{t_0}(y^2)\wt R_{20'}(x y)\wt R_{10'}(x^2) \right]=\nonumber\\
&\tr\left[ \left( \bar{K}_{0'}\left(\frac{1}{x}\right)R_{0'0}(x y)\bar{K}_{0}\left(\frac{1}{y}\right)
P_{02}P_{0'1}\right)\left( \wt R_{20}(y^2)\wt R_{10}(x y)\wt R_{20'}(x y)\wt R_{10'}(x^2)\right)\right],\label{A2}
\end{align}
where we defined $\tr\stackrel{\text{def}}{=}\tr_{0,0^{\prime}}$.

Multiplying \eqref{A2} by $R_{21}\left({\ds\frac{y}{x}}\right)$ from the right and
applying (\ref{almost Y-B}) twice, we bring the LHS of \eqref{YBE8}
to the form
\begin{align}
&\text{LHS}=\nonumber\\
&\tr\left[\bar{K}_{0'}\left(\frac{1}{x}\right) R_{0'0}(x y)\bar{K}_{0}\left(\frac{1}{y}\right)P_{02}P_{0'1}
R_{21}\left(\frac{y}{x}\right)\wt R_{10}(x y)\wt R_{20}(y^2)\wt R_{10'}(x^2) \wt R_{20'}(x y)\right]=\nonumber\\
&\tr\left[R_{0'0}\left(\dfrac{y}{x}\right) \bar{K}_{0}\left(\dfrac{1}{y}\right) R_{00'}(x y)\bar{K}_{0'}
\left(\dfrac{1}{x}\right)P_{02}P_{0'1}\wt R_{10}(x y)\wt R_{20}(y^2)\wt R_{10'}(x^2) \wt R_{20'}(x y) \right]=\nonumber\\
& \tr\left[\bar{K}_{0}\left(\dfrac{1}{y}\right) R_{00'}(x y)\bar{K}_{0'}\left(\dfrac{1}{x}\right)P_{02}P_{0'1}
\wt R_{10}(x y)\wt R_{20}(y^2)\wt R_{10'}(x^2) \wt R_{20'}(x y) R_{0'0}\left(\dfrac{y}{x}\right) \right],\label{A3}
\end{align}
where in the second line we used the reflection equation (\ref{YBE10}).

Repeating the above calculations for the RHS of the equation (\ref{YBE8}), we obtain
\bea
&\text{RHS}=\nonumber\\&\tr\left[\bar{K}_{0}\left(\dfrac{1}{y}\right) R_{00'}(x y)\bar{K}_{0'}
         \left(\dfrac{1}{x}\right)P_{02}P_{0'1}R_{0'0}\left(\dfrac{y}{x}\right)\wt R_{10'}(x^2)
         \wt R_{10}(x y) \wt R_{20'}(x y)\wt R_{20}(y^2) \right].\label{A4}
\end{align}
Applying \eqref{almost Y-B2} twice to \eqref{A4}, we bring it to \eqref{A3} which completes the proof.

Now let us show that the inverse mapping  from $\wt{K}(x)$ to $\bar{K}(x)$
has the form \eqref{YBE12}
\beq\label{A5}
    \bar{K}_1(x)=\op{tr}_{0'}\left[\wt{K}_{0'}\left(\dfrac{1}{x}\right)R_{0'1}\left(\dfrac{1}{x^2}\right) P_{0'1}\right].
\eeq
The proof of this statement is based on the trace property
\beq\label{A6}
   \op{tr}_{0'} \left(P_{00'} A_{10}B_{10'}\right)=\left(A_{10}^{t_0}B_{10}^{t_0} \right)^{t_0},
\eeq
which is valid for any matrices $A$ and $B$. Then the following calculation proves \eqref{A5}:
\bea
  &\op{tr}_{0'}\left[\wt{K}_{0'}\left(\dfrac{1}{x}\right)R_{0'1}\left(\dfrac{1}{x^2}\right) P_{0'1}\right]=
  \tr_{0,0^{\prime}}\left[\bar{K}_0\left(x\right)\wt R_{00'}\left(\dfrac{1}{x^2}\right) P_{00'}R_{0'1}
  \left(\dfrac{1}{x^2}\right) P_{0'1}\right]\nonumber\\&=\op{tr}_{0}\left[P_{01}\bar{K}_1\left(x\right)\op{tr}_{0'}
  \left\{P_{00'}\wt R_{10}\left(\dfrac{1}{x^2}\right)R_{10'}\left(\dfrac{1}{x^2}\right) \right\} \right]\nonumber\\
  &=\op{tr}_{0}\left[P_{01}\bar{K}_1\left(x\right) \left(\wt R_{10}^{t_0}\left(\dfrac{1}{x^2}\right)R_{10}^{t_0}
  \left(\dfrac{1}{x^2}\right) \right)^{t_0} \right]=\bar{K}_1(x).\label{A7}
\end{align}

\section {Useful summation formulas}\label{Basic summation formulas}\label{AppB}
First, we list some well known summation formulas from \cite{gasper}.\\

\noindent The $q$-binomial theorem:
\beq\label{q-binomial theorem}
    { }_1 \phi_0\left(q^{-n} ;-; q, z\right)=\left(z q^{-n} ; q\right)_n;
\eeq
the $q$-Vandermonde summation formula:
\beq\label{q-Vandermonde}
    { }_2 \phi_1\left(a, q^{-n} ; c ; q, q\right)=\frac{(c / a ; q)_n}{(c ; q)_n} a^n;
\eeq
the $q$-Pfaff-Saalschutz summation formula:
\beq\label{Pfaff-Saalschutz sum}
    { }_3 \phi_2\left[\bear{c}
a, b, q^{-n} \\
c, a b c^{-1} q^{1-n}
\eear ; q, q\right]=\frac{(c / a, c / b ; q)_n}{(c, c / a b ; q)_n}.
\eeq
We also give two special  $m$-dimensional extensions of the $q$-Pfaff-Saalschutz sum:
  \bea\label{Pfaff-Saalschutz sum (m dim)_2}
    \sum_{k_1, \ldots, k_m=0} q^{|\bd{k}|}
    \frac{(q^{-n}, b ; q)_{|\bd{k}|}}{\left(c, b c^{-1} q^{1-n}\prod\limits_{i=1}^m z_i; q\right)_{|\bd{k}|}}
    \prod\limits_{j=1}^{m}
     \frac{\left(z_j;q\right)_{k_j}}{(q ; q)_{k_j}}z_j^{\sum\limits_{i<j} k_i}=
    \frac{\left(c\prod\limits_{i=1}^m z_i^{-1}, \dfrac{c}{b}; q\right)_{n}}{\left(c,
    \dfrac{c}{b}\prod\limits_{i=1}^m z_i^{-1} ; q\right)_{n}},
\end{align}
  \beq\label{Pfaff-Saalschutz sum (m dim)_1}
    \sum_{k_1, \ldots, k_m=0} q^{|\bd{k}|-\sum\limits_{i<j} k_i n_j}
    \frac{(a, b ; q)_{|\bd{k}|}}{\left(c, a b c^{-1} q^{1-|n|} ; q\right)_{|\bd{k}|}}
    \prod_{i=1}^m \frac{\left(q^{-n_i} ; q\right)_{k_i}}{(q ; q)_{k_i}}=
    \frac{(c / a, c / b ; q)_{|\bd{n}|}}{(c, c / a b ; q)_{|\bd{n}|}}.
\eeq
Both relations can be proven as follows. First, we rewrite the summations in the form
$\sum\limits_{k_1, \ldots, k_m=0}\rightarrow \sum\limits_{l=0}\sum\limits_{|k|=l}$. Then the result follows
from (\ref{Pfaff-Saalschutz sum}) and the formula
\beq\label{summation formula}
    \sum_{|k|=l} \prod\limits_{j=1}^{m}z_j^{\sum\limits_{i<j} k_i} \prod_{i=1}^m
    \frac{\left(z_i ; q\right)_{k_i}}{(q ; q)_{k_i}}=\frac{\left(\prod\limits_{i=1}^m z_i ; q\right)_l}{(q ; q)_l},
\eeq
where the last equality can be easily proved by induction on $m$ and using the $q$-Vandermonde summation
formula (\ref{q-Vandermonde}).

\section {Proof of (\ref{first sum of V})}\label{AppC}
The proof of the \eqref{first sum of V} is based on  (\ref{Pfaff-Saalschutz sum (m dim)_2}).
First, we shift the summation index $\bd{d}\rightarrow \bd{d}+\bd{c}$ and use
the definition of the function $V$  (\ref{V-function}) to get
\beq
    \bed
        &\sum\limits_{\bd{d}}q^{-2\sum\limits_{k=1}^{n-1}k(d_k-c_k)+n(|\bd{d}|-|\bd{c}|)}\frac{V_{q^{-J}}(\bd{d},\bd{b})
        V_{y}(\bd{d},\bd{c})}{V_{z}(\bd{d},\bd{b})V_{\frac{y}{z} q^{n-J}}(\bd{d},\bd{m})}=\\&\frac{(q^{n-1}y)^{|\bd{c}|-|\bd{m}|}}
        {(q^J z)^{|\bd{b}|-|\bd{m}|}} \frac{q^{Q(\bd{c},\bd{c}-\bd{m})}}{q^{Q(\bd{c}-\bd{m},\bd{m})}}
        \frac{\prod\limits_{s=1}^{n-1}\left(q^2;q^2\right)_{c_s-m_s} \left(q^{-2
   J};q^2\right)_{|\bd{c}|-|\bd{b}|}}{\left(\frac{q^{2n-2J} y^2}{z^2};q^2\right)_{|\bd{c}|-|\bd{m}|}
   \left(z^2;q^2\right)_{|\bd{c}|-|\bd{b}|}}\times\\&\sum_{\bd{d}}\frac{\left(\frac{q^{2 (|\bd{c}| -|\bd{b}|)}}{q^{2 J}},
   y^2;q^2\right)_{|\bd{d}|}q^{2\sum\limits_{k=1}^{n}(n-k)d_k +2Q(\bd{d},\bd{c}-\bd{m})}}{\left(\frac{q^{-2 (J-n-(|\bd{c}| -|\bd{m}| ))}
   y^2}{z^2}, q^{2 (|\bd{c}| -|\bd{b}|)} z^2;q^2\right)_{|\bd{d}|}}
   \prod_{s=1}^{n-1}\frac{\left(q^{2 \left(1+c_s-m_s\right)};q^2\right)_{d_s}}{\left(q^2;q^2\right)_{d_s}}.
    \eed
\eeq
After taking into account the relation
\beq
    q^{2\sum\limits_{k=1}^{n}(n-k)d_k +2Q(\bd{d},\bd{c}-\bd{m})}=q^{2|\bd{d}|}
    \prod_{j=1}^{n-1}q^{2\sum\limits_{i<j} \left(1+c_j-m_j\right)d_i},
\eeq
we can apply  (\ref{Pfaff-Saalschutz sum (m dim)_2})
\beq
    \bed
      &\sum\limits_{\bd{d}}q^{-2\sum\limits_{k=1}^{n-1}k(d_k-c_k)+n(|\bd{d}|-|\bd{c}|)}
      \frac{V_{q^{-J}}(\bd{d},\bd{b}) V_{y}(\bd{d},\bd{c})}{V_{z}(\bd{d},\bd{b})
      V_{\frac{y}{z} q^{n-J}}(\bd{d},\bd{m})}=\frac{(q^{n-1}y)^{|\bd{c}|-|\bd{m}|}}
      {(q^J z)^{|\bd{b}|-|\bd{m}|}} \frac{q^{Q(\bd{c},\bd{c}-\bd{m})}}{q^{Q(\bd{c}-\bd{m},\bd{m})}}
      \times\\&\frac{\prod\limits_{s=1}^{n-1}\left(q^2;q^2\right)_{c_s-m_s} \left(q^{-2
   J};q^2\right)_{|\bd{c}|-|\bd{b}|}}{\left(\frac{q^{2n-2J} y^2}{z^2};q^2\right)_{|\bd{c}|-|\bd{m}|}
   \left(z^2;q^2\right)_{|\bd{c}|-|\bd{b}|}} \frac{\left(\frac{q^{2-2 J} y^2}{z^2},
   \frac{q^{-2 (J-n-| c| +| m| )}}{z^2};q^2\right)_{J-(|\bd{c}|-|\bd{b}|)}}{\left(\frac{q^{-2 (J-n-| c| +| m| )}
   y^2}{z^2},\frac{q^{2-2 J}}{z^2};q^2\right)_{J-(|\bd{c}|-|\bd{b}|)}}.
    \eed
\eeq
Then successively applying the following identities:
\beq
    \left(a q^{-2J};q^2\right)_{J-(|\bd{c}| -|\bd{b}|)}=\dfrac{\left(q^2/a;q^2\right)_J}
    {\left(q^2/a;q^2\right)_{|\bd{c}| -|\bd{b}|}}\left(-\dfrac{a}{q^2}\right)^{J-(|\bd{c}| -|\bd{b}|)}
     \frac{q^{(|\bd{c}| -|\bd{b}|) (|\bd{c}| -|\bd{b}|-1)}}{q^{J(J-1)}},
\eeq
\beq
    \left(a q^{-2 (|\bd{c}| -|\bd{m}| )};q^2\right)_J=\dfrac{\left(a;q^2\right)_J
    \left(q^2/a;q^2\right)_{|\bd{c}| -|\bd{m}| }}
    {\left(q^{2 (1-J)}/a;q^2\right)_{|\bd{c}| -|\bd{m}|}} q^{-2 J (|\bd{c}| -|\bd{m}|)},
\eeq
we see, that the factors of the form $\left(a;q^2\right)_{J}$ are combined in
the function $\mu_J^{(n)}(y,z)$ from (\ref{mu function}), and one obtains
\beq\label{lhs after all transfrom}
     \bed
          &\text{LHS (\ref{first sum of V})}=\mu_J^{(n)}(y,z)\frac{(q^{n-1}y)^{|\bd{c}|-|\bd{m}|}}
          {(q^J z)^{|\bd{b}|-|\bd{m}|}} \frac{q^{Q(\bd{c},\bd{c}-\bd{m})}}{q^{Q(\bd{c}-\bd{m},\bd{m})}}
          \dfrac{\prod\limits_{s=1}^{n-1}\left(q^2;q^2\right)_{c_s-m_s}}{\left(\frac{q^{2 n} y^2}
          {z^2};q^2\right)_{|\bd{c}| -|\bd{m}|}}\times \\ &\dfrac{\left(q^{-2 J};
          q^2\right)_{|\bd{c}| -|\bd{b}| }}{\left(\frac{z^2}{y^2};q^2\right)_{|\bd{c}| -|\bd{b}|}}
          \dfrac{\left(\frac{q^{2 n}}{z^2};q^2\right)_{|\bd{c}| -|\bd{m}| }}{\left(\frac{q^{-2 J+2 n}}
          {z^2};q^2\right)_{|\bd{c}| -|\bd{m}| }}\dfrac{\left(\frac{q^{-2 (-1+n+|\bd{c}| -|\bd{m}| )} z^2}
          {y^2};q^2\right)_{|\bd{c}| -|\bd{b}| }}
          {\left(q^{-2 (-1+n+|\bd{c}| -|\bd{m}| )} z^2;q^2\right)_{|\bd{c}| -|\bd{b}| }}.
     \eed
\eeq
Substituting to  the RHS of \eqref{first sum of V} the definition (\ref{V-function}) of
 the function  $V_x(\bd{a},\bd{b})$, we come to the same expression
\eqref{lhs after all transfrom}.

  \section{Some explicit matrices}\label{Some explicit matrices}\label{AppD}
  Here we demonstrate the structure of some triangular boundary matrices. 
  For the six-vertex case $n=2$ and $J=1$ 
  we obtain from (\ref{Upper-triangular solution}),
  (\ref{K-matrix, lower-triangular solution}), (\ref{Left boundary}) and (\ref{Left boundary, lower-triangular solution})
  \beq\label{Kn21}
  \bed K^{(u)}_1(y)=\left(
  \bear{cc}
 1& \frac{q \nu(1-y^4)}{y^2(1-q \nu y^2)}\\
  0& \frac{y^2-q \nu}{y^2(1-q \nu y^2)}\\
  \eear\right),\quad
  K^{(d)}_1(y)=\left(
  \bear{cc}
\frac{y^2(y^2-q \nu)}{1-q \nu y^2_{\phantom{I}}}&0\\
  \frac{1-y^4}{1-q \nu y^2}& 1\\
  \eear\right),
  \eed\eeq
  \beq\label{Kn22}
  \bed \bar K^{(u)}_1(y)=\left(
  \bear{cc}
 1& \frac{y^4-1}{y^2(y^2-q\nu)}\\
  0&\frac{1-q\nu y^2}{y^2(y^2-q \nu ) }\\
  \eear\right),\quad
  \bar K^{(d)}_1(y)=\left(
  \bear{cc}
\frac{y^2(1-q \nu y^2)}{y^2-q\nu}&0\\
  \frac{q\nu(y^4-1)}{y^2-q\nu}& 1\\
  \eear\right).
  \eed\eeq 
 For $J=1$ and
$n>2$ we refer to \cite{Crampe14,Ragoucy2016}.
  
We shall also present the simplest case  
 of the higher spin matrices for two particles, i.e when $J=2$ and $n=3$.
  The basis is ordered as follows:
  \beq
  |0,0\rangle, \quad |0,1\rangle, \quad |0,2\rangle, \quad |1,0\rangle, \quad |1,1\rangle, \quad |2,0\rangle.
  \eeq
Let us introduce the parameters
  \beq
      \la=\frac{y^2}{\nu  q^2}, \quad  \mu=\frac{1}{y^2 \nu  q^2}, \quad p=q^2.
  \eeq
  Then the upper- and lower-triangular matrices (\ref{Upper-triangular solution}),
  (\ref{K-matrix, lower-triangular solution}), describing the right boundary, have the form
   \beq
   \bed
  &K^{(u)}_2(y)=\\&\left(
\bear{cccccc}
 1 & \frac{\left(\frac{\mu}{\la};p\right)_1}{\left(\mu;p\right)_1} &
 \frac{\left(\frac{\mu}{\la};p\right)_2}{\left(\mu;p\right)_2} &
   \frac{\left(\frac{\mu}{\la};p\right)_1}{\left(\mu;p\right)_1} &
   \frac{\left(\frac{\mu}{\la};p\right)_2}{\left(\mu;p\right)_2} &
   \frac{\left(\frac{\mu}{\la};p\right)_2}{\left(\mu;p\right)_2} \\
 0 & \frac{\mu}{\la}\frac{\left(\la;p\right)_1}{\left(\mu;p\right)_1} &
 \frac{\mu}{\la}\frac{\left(\la,\frac{\mu}{\la};p\right)_1}{\left(\mu;p\right)_2}
 {2\brack 1}_{p} & 0 & p\frac{\mu}{\la}\frac{\left(\la,\frac{\mu}{\la};p\right)_1}{\left(\mu;p\right)_2} & 0 \\
 0 & 0 & \left(\frac{\mu}{\la}\right)^2\frac{\left(\la;p\right)_2}{\left(\mu;p\right)_2} & 0 & 0 & 0 \\
 0 & 0 & 0 &  \frac{\mu}{\la}\frac{\left(\la;p\right)_1}{\left(\mu;p\right)_1} &
  \frac{\mu}{\la}\frac{\left(\la,\frac{\mu}{\la};p\right)_1}{\left(\mu;p\right)_2} &
    \frac{\mu}{\la}\frac{\left(\la,\frac{\mu}{\la};p\right)_1}{\left(\mu;p\right)_2}{2\brack 1}_{p} \\
 0 & 0 & 0 & 0 &  \left(\frac{\mu}{\la}\right)^2\frac{\left(\la;p\right)_2}{\left(\mu;p\right)_2} & 0 \\
 0 & 0 & 0 & 0 & 0 & \left(\frac{\mu}{\la}\right)^2\frac{\left(\la;p\right)_2}{\left(\mu;p\right)_2} \\
\eear
\right),
\eed
\eeq
\beq
   \bed
  &K^{(d)}_2(y)=\\&\left(
\bear{cccccc}
 \frac{\left(\la ;p\right)_2}{\left(\mu ;p\right)_2} & 0 & 0 & 0 & 0 & 0 \\
 0 & \frac{\left(\la ;p\right)_2}{\left(\mu ;p\right)_2} & 0 & 0 & 0 & 0 \\
 0 & 0 & \frac{\left(\la ;p\right)_2}{\left(\mu ;p\right)_2} & 0 & 0 & 0 \\
 \la \frac{\left(\la,\frac{\mu}{\la};p\right)_1}{\left(\mu ;p\right)_2}{2\brack 1}_{p}
 & p\la\frac{\left(\la,\frac{\mu }{\la };p\right)_1}{\left(\mu ;p\right)_2} & 0 &
 \frac{\left(\la ;p\right)_1}{\left(\mu ;p\right)_1}
   & 0 & 0 \\
 0 & \la\frac{\left(\la, \frac{\mu }{\la };p\right)_1}{\left(\mu ;p\right)_2} &
  \la\frac{\left(\la,\frac{\mu }{\la };p\right)_1}{\left(\mu ;p\right)_2}{2\brack 1}_{p}
   & 0 & \frac{\left(\la ;p\right)_1}{\left(\mu ;p\right)_1} & 0 \\
 \la ^2\frac{\left(\frac{\mu }{\la };p\right)_2}{\left(\mu ;p\right)_2} &
  \la ^2\frac{\left(\frac{\mu }{\la };p\right)_2}{\left(\mu ;p\right)_2} &
   \la ^2\frac{\left(\frac{\mu }{\la };p\right)_2}{\left(\mu ;p\right)_2} &
   \la\frac{ \left(\frac{\mu }{\la };p\right)_1}{\left(\mu ;p\right)_1} &
   \la \frac{\left(\frac{\mu }{\la };p\right)_1}{\left(\mu ;p\right)_1} & 1 \\
\eear
\right).
\eed
\eeq
The solutions (\ref{Left boundary}), (\ref{Left boundary, lower-triangular solution}),
related to the left boundary, have a similar structure:
\beq
\bed
   & \bar{K}^{(u)}_2(y)=\\&\left(
\bear{cccccc}
 1 & \mu\frac{\left(\frac{\la }{\mu };p\right)_1}{\left(\la ;p\right)_1} &
 \mu^2\frac{\left(\frac{\la }{\mu };p\right)_2}{\left(\la ;p\right)_2} &
   \mu \frac{\left(\frac{\la }{\mu };p\right)_1}{\left(\la ;p\right)_1} &
   \mu^2\frac{\left(\frac{\la }{\mu };p\right)_2}{\left(\la ;p\right)_2} &
   \mu^2\frac{\left(\frac{\la }{\mu };p\right)_2}{\left(\la ;p\right)_2} \\
 0 & \frac{\left(\mu ;p\right)_1}{\left(\la ;p\right)_1} & \mu\frac{\left(\mu,
 \frac{\la }{\mu };p\right)_1}{\left(\la ;p\right)_2}{2\brack 1}_{p} & 0 &
 \mu\frac{\left(\mu, \frac{\la }{\mu };p\right)_1}{\left(\la ;p\right)_2} &
   0 \\
 0 & 0 & \frac{\left(\mu ;p\right)_2}{\left(\la ;p\right)_2} & 0 & 0 & 0 \\
 0 & 0 & 0 & \frac{\left(\mu ;p\right)_1}{\left(\la ;p\right)_1} & \mu  p
 \frac{\left(\mu ,\frac{\la }{\mu};p\right)_1}{\left(\la ;p\right)_2} &
 \mu\frac{\left(\mu,\frac{\la }{\mu };p\right)_1}{\left(\la ;p\right)_2}{2\brack 1}_{p} \\
 0 & 0 & 0 & 0 & \frac{\left(\mu ;p\right)_2}{\left(\la ;p\right)_2} & 0 \\
 0 & 0 & 0 & 0 & 0 & \frac{\left(\mu ;p\right)_2}{\left(\la ;p\right)_2} \\
\eear
\right),
\eed
\eeq
\beq
\bed
    &\bar{K}^{(d)}_2(y)=\\&\left(
\bear{cccccc}
 \left(\frac{\la }{\mu }\right)^2 \frac{\left(\mu ;p\right)_2}{\left(\la ;p\right)_2} & 0 & 0 & 0 & 0 & 0 \\
 0 & \left(\frac{\la }{\mu }\right)^2 \frac{\left(\mu ;p\right)_2}{\left(\la ;p\right)_2} & 0 & 0 & 0 & 0 \\
 0 & 0 & \left(\frac{\la }{\mu }\right)^2 \frac{\left(\mu ;p\right)_2}{\left(\la ;p\right)_2} & 0 & 0 & 0 \\
 \frac{\la }{\mu }\frac{\left(\mu, \frac{\la }{\mu };p\right)_1}
 {\left(\la ;p\right)_2}{2\brack 1}_{p} &  \frac{\la }{\mu }\frac{\left(\mu, \frac{\la }
 {\mu };p\right)_1}{\left(\la ;p\right)_2} & 0 &  \frac{\la }{\mu }\frac{\left(\mu ;p\right)_1}{\left(\la
   ;p\right)_1} & 0 & 0 \\
 0 & p\frac{\la }{\mu }\frac{\left(\mu, \frac{\la }{\mu };p\right)_1}{\left(\la ;p\right)_2} &
 \frac{\la }{\mu }\frac{\left(\mu, \frac{\la }{\mu };p\right)_1}{\left(\la ;p\right)_2}
 {2\brack 1}_{p} & 0 &  \frac{\la }{\mu }\frac{\left(\mu ;p\right)_1}{\left(\la ;p\right)_1} & 0 \\
 \frac{\left(\frac{\la }{\mu };p\right)_2}{\left(\la ;p\right)_2} &
 \frac{\left(\frac{\la }{\mu };p\right)_2}{\left(\la ;p\right)_2} &
   \frac{\left(\frac{\la }{\mu };p\right)_2}{\left(\la ;p\right)_2} &
    \frac{\left(\frac{\la }{\mu };p\right)_1}{\left(\la ;p\right)_1} &
   \frac{\left(\frac{\la }{\mu };p\right)_1}{\left(\la ;p\right)_1} & 1 \\
\eear
\right).
\eed
\eeq
\\
\\


\end{document}